\documentclass[aps,prl,twocolumn,superscriptaddress,groupedaddress,floatfix,final]{revtex4} 
\usepackage{graphicx}  
\usepackage{dcolumn}   
\usepackage{bm}        
\usepackage{amssymb}   
\usepackage{amsmath}
\usepackage{mathtools}
\usepackage{float}
\usepackage{color}
\usepackage{soul}
\usepackage{subfiles}
\usepackage[normalem]{ulem}
\begin{document}



\title{Relaxation of a Mott-neuron}

\author{Federico Tesler}

\affiliation{Departamento de F\'{i}sica, Facultad de Ciencias Exactas y Naturales, Universidad de Buenos Aires and IFIBA, CONICET, Cuidad Universitaria, Buenos Aires 1428, Argentina}

\author{Coline Adda}

\affiliation{Institut des Mat\'{e}riaux Jean Rouxel (IMN), Universit\'{e} de Nantes, CNRS,2 rue de la Houssini\`{e}re, BP 32229,  44322 Nantes Cedex 3, France}

\author{Julien Tranchant}

\affiliation{Institut des Mat\'{e}riaux Jean Rouxel (IMN), Universit\'{e} de Nantes, CNRS,2 rue de la Houssini\`{e}re, BP 32229,  44322 Nantes Cedex 3, France}

\author{Benoit Corraze}

\affiliation{Institut des Mat\'{e}riaux Jean Rouxel (IMN), Universit\'{e} de Nantes, CNRS,2 rue de la Houssini\`{e}re, BP 32229,  44322 Nantes Cedex 3, France}

\author{Etienne Janod}

\affiliation{Institut des Mat\'{e}riaux Jean Rouxel (IMN), Universit\'{e} de Nantes, CNRS,2 rue de la Houssini\`{e}re, BP 32229,  44322 Nantes Cedex 3, France}

\author{Laurent Cario}

\affiliation{Institut des Mat\'{e}riaux Jean Rouxel (IMN), Universit\'{e} de Nantes, CNRS,2 rue de la Houssini\`{e}re, BP 32229,  44322 Nantes Cedex 3, France}

\author{Pablo Stoliar}

\affiliation{CIC nanoGUNE, Tolosa Hiribidea 76,  20018 Donostia-San Sebastian, Spain}

\author{Marcelo Rozenberg}

\affiliation{Laboratoire de Physique des Solides, CNRS, Univ. Paris-Sud, Universit\'{e} Paris-Saclay, 91405 Orsay Cedex, France}

\affiliation{Physics Department, University of California-San Diego, La Jolla California 92093-0319, USA}


\begin{abstract}
We consider the phenomenon of electric Mott transition (EMT), which is an electric induced
insulator to metal transition. Experimentally, it is observed that depending on the magnitude
of the electric excitation the final state may show a short lived or a long lived resistance change.
We extend a previous model for the EMT to include the effect of local structural distortions 
through an elastic energy term.
We find that by strong electric pulsing the induced metastable phase may become
further stabilized by the electro-elastic effect.
We present a systematic study of the model by numerical simulations and compare the results to new experiments in Mott insulators of the AM$_4$Q$_8$ family. Our work significantly extends the scope of our recently introduced leaky-integrate-and-fire Mott-neuron [P. Stoliar Adv Mat 2017] to bring new insight on the physical mechanism of its relaxation. This is a key feature for future neuromorphic circuit implementations.

\end{abstract}

\maketitle


\subsection{Introduction}

The information age we live in is supported on a physical under-layer of electronic hardware, which originates
in condensed matter physics research. The mighty progress made in recent decades produced faster and
more power efficient electronic devices. This enabled a seemingly endless improvement in computer performance.
However, with the smallest feature size of transistors reaching down to mere 5 nm, the technology is reaching 
an unavoidable physical limit.
This situation concerns the current computational paradigm, based on binary logic and the von Neumann architecture.
While incremental progress along this direction is likely to go on, the search for disruptive technologies is on.

On the other hand, we are currently witnessing significant progress in artificial intelligence. In fact, older algorithmic 
paradigms, such as neural networks combined with newer ideas such as wavelet based filtering have produced a 
remarkable improvement of performance of computers at dealing with tasks where they were traditionally poor, such as pattern
recognition. This area of research is known as Deep Neural Networks\cite{Mallat2012, Schmidhuber201585}. However, most of this progress has been
reached by running the new algorithms with conventional codes on conventional (i.e. von Neumann) computers.
While the performance of these algorithms is remarkable at many tasks, unfortunately, it is
limited by the computational power of current machines. For instance, the number of 
synapses in a brain is 10$^{15}$, thus even if  one allocates a single memory per synapse, 
conventional computers fall irreparably short.

Therefore, an exciting perspective is to implement the neural networks directly on hardware, by realizing novel
electronic devices that may directly implement the neural network functionality.
These bio-inspired electronic circuits have two key new components. 
One is an ``synapse analogue'' and the other is an ``neuron analogue''. 
The former can be implemented by a non-volatile resistance, whose value can be programmed (and re-programmed)
by application of voltage pulses. This can be effectively done by the so called RRAM devices (for resistive random access memories) 
and also known as memristors (for memory dependent resistors). One widely adopted implementation of these systems is
in the form of capacitor-like structures, where the two electrodes are ordinary metals (Pt, Ti, etc) and the dielectric is a transition metal 
oxide \cite{Rozenberg2011}. The resistive change is observed in the two-terminal resistance of the device, which can be modified
by electric pulsing \cite{Janod2015}. These devices have been intensively investigated and developed 
during the past decade.
The basic physical mechanisms for the resistive change  in these oxide memristive devices are now understood in 
certain detail \cite{Wong2012}. 
They are mainly based on achieving structural changes by inducing oxygen or metallic ion migration \cite{Shao}.
After more than a decade of intense research these type of devices are beginning to reach the market.

This significant progress enables the implementation of ``synapses analogue'' \cite{Synapse1,Pershin2010881}, 
unfortunately, the progress in the research and 
implementation of ``neurons analogue'' has not been comparable. Hence, their development becomes urgent.

In the context of resistive change systems we may mention two important contributions towards devices with 
a ``neuron'' inspired functionality: one is the ``neuristor'', which realizes a functionality similar to the Hodgkin-Huxley model 
for electric pulse propagation in biological axons \cite{pickett2013}. 
The other one is the ``artificial neuron'', which implements the functionality of a leaky-integrate-and-fire (LIF) 
biological spiking neuron \cite{Stoliar2017}. Remarkably, both implementations are based on Mott insulators, which are
strongly correlated systems that exhibit metal-insulator transitions \cite{RMP-IFT}. Mott systems have been intensively
investigated since the 80's, following the discovery of high temperature superconductivity in cuprates. Their basic feature of
Mott insulators is that according to their expected band-structure they should be metals, with partially filled bands, 
however, they turn out to be insulators. The physical reason is due to the effect of the strong local Coulomb
repulsive interaction between electrons in $d$-orbital bands that localize the electrons \cite{RMP-IFT}.

Here we shall focus on the latter of those two types of artificial Mott neuron systems. More specifically,
we shall gain insight in a key feature of the LIF model, namely the ``leaky'' property and shall also extend our model to study
the recovery behavior of the Mott neuron right after the firing event which is corresponding to the ``refractory period'' of a
biological neuron.
In the LIF model, a neuron is excited by a train of incoming electric pulses (spikes), which are {\it integrated}
in time by the neuron, until it reaches a threshold level where it {\it fires} a spike. The ``fire'' event in the Mott neuron corresponds to the EMT, that is the resistance collapse of the Mott insulator.
The {\it leaky} property models the fact that in the ``down'' time in-between pulses
the excitation level of the neuron slowly decays with a given relaxation time constant. 
Evidently, for a neuron to reach a threshold and fire, the time-delay between arriving pulses 
has to be shorter than the relaxation time constant of that neuron  \cite{Stoliar2013}.
Therefore, the control of the ``leaky time'' is a key issue. On the other hand, the ``refractory time''  of a biological neuron 
refers to the period of time {\it right after} the fire event while the neuron cannot fire again. 
As we shall see below, in our artificial Mott neuron model both features are related to the same underlying mechanism,
which leads to two characteristic relaxation times of the system. 

One may naively expect that the relaxation time would be an intrinsic constant of the material, 
however the situation is more complex
since the artificial neurons are based on strongly correlated Mott materials that undergo insulator-metal phase transitions. 
In systems in equilibrium, Mott transitions can be induced and controlled by a variety of parameters, such as
pressure, chemical doping and temperature. However, in the present situation, the understanding of the relaxation is
further complicated by the fact that the LIF functionality is a property of the system out-of-equilibrium. 
That is, when the Electric Mott Transition (EMT) is induced by pulses of electric field that
provoke a partial dielectric breakdown. During this breakdown, the resistance collapses \cite{Guiot2013, Stoliar2013}, 
and eventually recovers back to the original value after the application of the electric pulse is terminated. The theoretical
understanding of the recovery of a Mott insulator from the collapsed resistance state is a challenging many-body
out of equilibrium problem of actual interest \cite{Aoki2014}. 

Experiments of EMT  show that the relaxation from the low to the high-resistance state of a Mott insulator 
depends significantly on the strength of the applied electric field \cite{Janod2015}. 
In fact, the systematic study of 
the field induced dielectric breakdown in the Mott system GaTa$_4$Se$_8$ and similar compounds
revealed the existence of three different regimes: 
(i) at low applied electric fields there is, of course, no large resistance change, 
(ii) at intermediate field strength there is a {\it volatile} change of resistance, that is, there is an initial sudden
collapse followed by a recovery. 
And (iii) at high fields there is a seemingly {\it non-volatile} change of resistance, that is, the resistance
collapses and does not recover or may take very long time (i.e., the retention time is virtually infinity).
In our artificial neuron model, we associate the duration of the resistance recovery to the ``refractory period'' 
mentioned above, and its characteristic time will be termed {\it retention time}.

We should note that the resistive change in regime (ii) is a perfectly reproducible effect and the breakdown 
occurs after a delay-time $t_D$ (of the order of $\mu$s to ms), strongly dependent on the applied voltage \cite{Janod2015}. 
When the applied voltage pulse is terminated, the low resistance relaxes back to the original high value in a characteristic
time that we shall denote $\tau_R$ \cite{Stoliar2017}.
In contrast, in regime (iii) the resistance change is permanent, i.e. $\tau_R \to \infty$, but the resistive changes are
more difficult to control and reproduce. 
We should also note that we are discussing the resistance and not the resistivity. It is still not fully understood how
the resistive transition occurs spatially. It is most likely through electronic filamentary structures formation \cite{Stoliar2013}

In previous works, we have introduced a model that accounts fairly well for the basic phenomenology
of the EMT \cite{Stoliar2013}. 
However, that model has a single relaxation time for the recovery of the ``broken-down'' metallic regions back to the
initial Mott insulator state. This feature clearly cannot account for the observation of the regime (iii) described above,
and neither to the behavior of $\tau_R$ with applied pulse strength in regime (ii) as is observed in the experiments that we shall report here. 
Thus, the main goal of the present work is to extend the previous model to capture the behavior of
the relaxation time in the experiments of the EMT.
We shall show that by introducing an electro-elastic effect, the relaxation time $\tau_R$ will display 
a non-trivial behavior. For instance, we shall show that the duration of the relaxation will depend 
on the volume fraction that has changed from Mott-insulator to metal during the application of the voltage pulses. 
Related to this and relevant for experiments, we shall see that by application of a higher voltage strength and/or a 
longer pulse duration one may induce the growth of {\it thicker} metallic filaments and achieve a 
significant increase of the relaxation time. 

\begin{figure}[h]
  \includegraphics[width=0.5\textwidth]{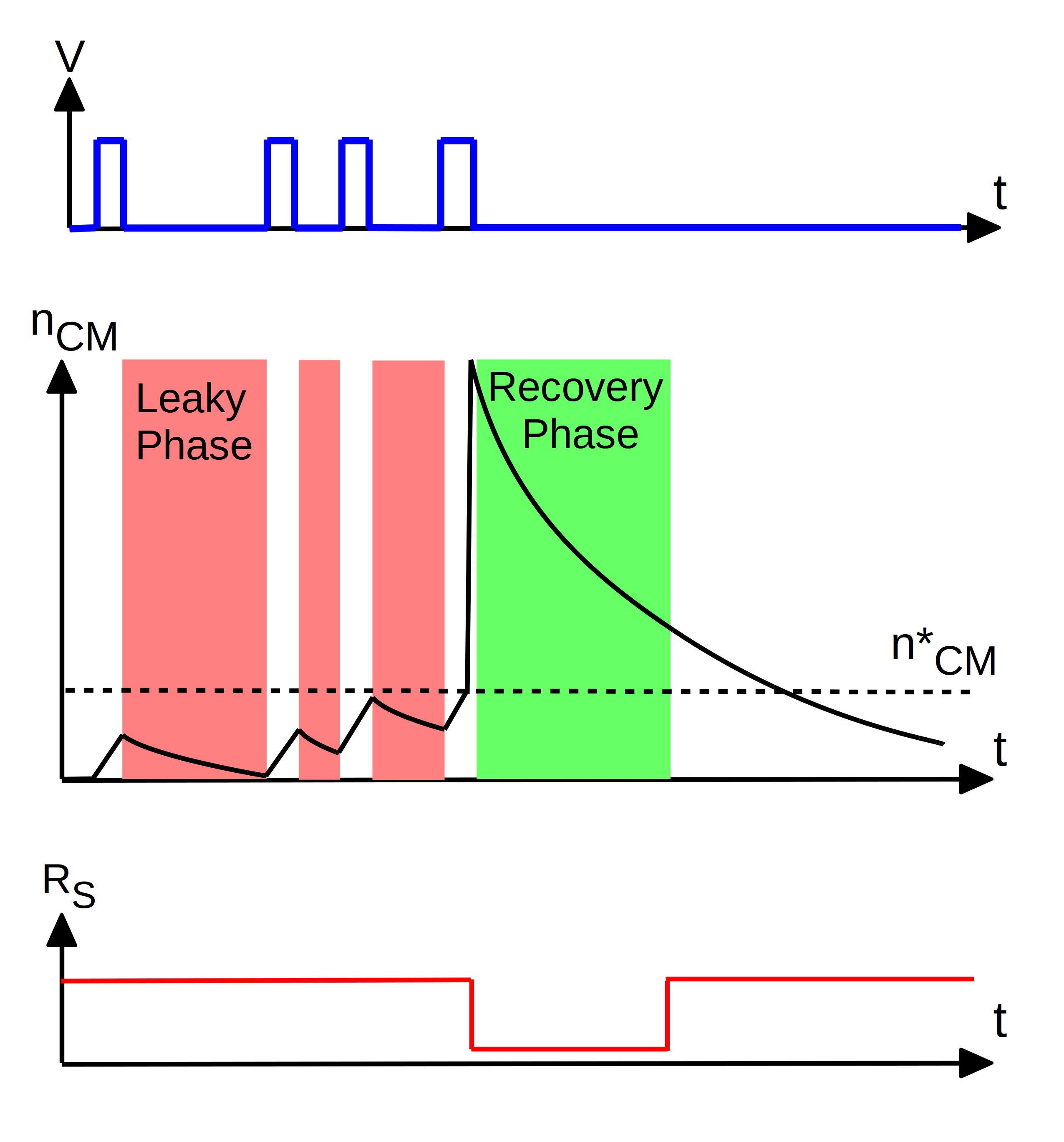}
  \caption{The leaky and refractory phases are respectively depicted in red and green.  
  The first take place in-between pulses and the second after the EMT, that is, after the neuron fires.   }
  \label{Fig_phases}
\end{figure}

Before introducing the model we should clarify a few important points, which connect the present study to a model of
a neuron analogue aimed at the implementation of bio-inspired neural networks. 
The resistive collapse is considered to result from a two-step
process. The initial one where the strong applied field enhances the rate of phase change and a larger number of
isolated small regions of the insulating material turn metallic. If the production rate is sufficiently large, more than the
rate of relaxation of those regions back to the insulator state, then the density of metallic regions will steadily increase.
Eventually it will reach a critical density where a sudden avalanche-like process will create a filament \cite{Stoliar2013}.
The resistance change during the first part of the process is quite small, and the collapse of its value is due to the
formation of the filament or, if the applied voltage is strong, of more filaments at random positions, further reducing the resistance.
That initial model assumed a single intrinsic time for the relaxation of metallic regions back to insulator state, irrespective
of whether it belonged to an isolated region or was part of a filamentary structure. This feature will be partially modified
in the model introduced in the present work. We shall make the additional assumption that the relaxation probability of a 
given cell depends on the current state of its neighboring cells. Hence, a metallic cell will effectively
have a different relaxation rate depending whether it is an isolated one or it is within a filamentary structure. 
The model will aim to describe the whole process of resistance collapse and recovery, namely the filament formation and its reabsorption. The {\it leaky-time} of a neuron analogue model \cite{Stoliar2017} is associated to the 
dynamics of the neuron in between arriving pulses {\it before} the fire. Correspondingly, in the context of the present model, 
the leaky behavior is associated to the behavior of the system during the initial filamentary formation process. 
More specifically, with the initial stage where the density of isolated metallic regions increase until the critical density 
is attained and with the ensuing rapid growth of the filamentary structure.
Experimentally, this leaky stage may be explored and characterized by the application of short electric pulses \cite{Stoliar2017}.
In contrast, there is a process of reabsorption of a formed filament, which corresponds to the resistance recovery from low to high, that begins when the pulse is terminated after the fire event (see Fig.1) This process is what one would associate to the ``refractory-time'' of a biological neuron. However, there is a significant difference between the refractory time of a biological neuron and the Mott neuron. In the former, the neuron is ``off'' during the refractory time, that is, arriving spikes cannot induce a new action potential. In the Mott neuron, in contrast, the system is in the low resistance state, which means that arriving pulses will generate current pulses. Because of this difference, we term the recovery of the high-resistance state (i.e. filament reabsorption and rupture) as ``recovery time'' instead of refractory time (see Fig. \ref{Fig_phases}).


As we shall see from the study of the simulation results of our model, the leaky process corresponds
to the relaxation of {\it isolated} metallic regions in an essentially insulating system, 
while the recovery period corresponds to the rupture or de-percolation process of already formed metallic filaments.
We shall see that both the leaky and the recovery phases are captured by our model. As we shall also
show in experiments that we report here, our modeling work provides useful
insight for the control of these relevant neuronal model features.

\subsection{The model} 

\begin{figure}[h]
  \includegraphics[width=0.5\textwidth]{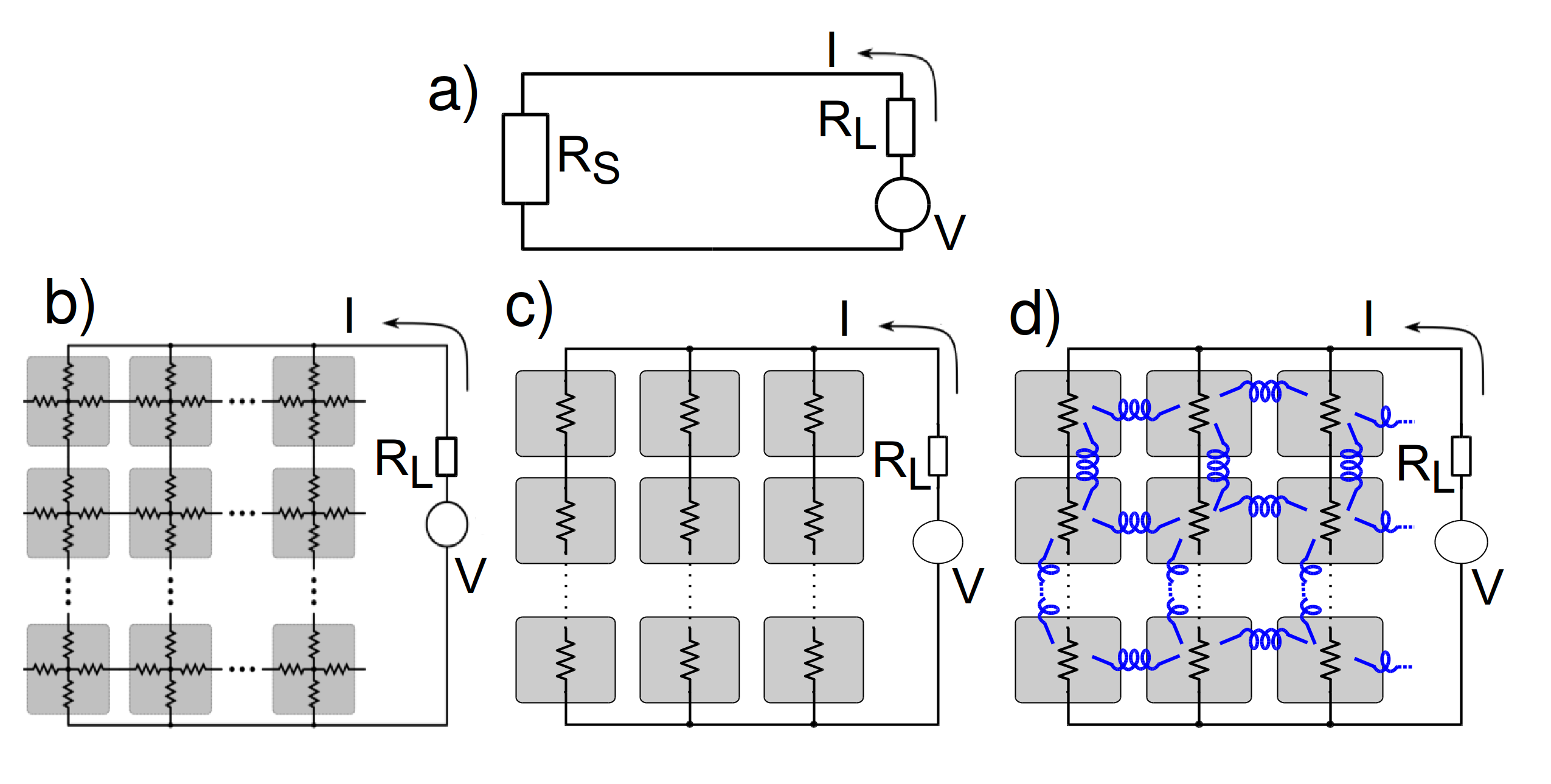}
  \caption{ 
  a) Circuit diagram of the system. b) Diagram of the 2D resistive network of the model from ref.\cite{Stoliar2013}. c) Diagram of the 1D filament approximation. All transverse currents have been neglected. d) Diagram of the 1D approximation with local elastic interactions (represented by the blue springs).
  }
   
  \label{Fig_intro}
\end{figure}

\begin{figure}[h]
  \includegraphics[width=0.5\textwidth]{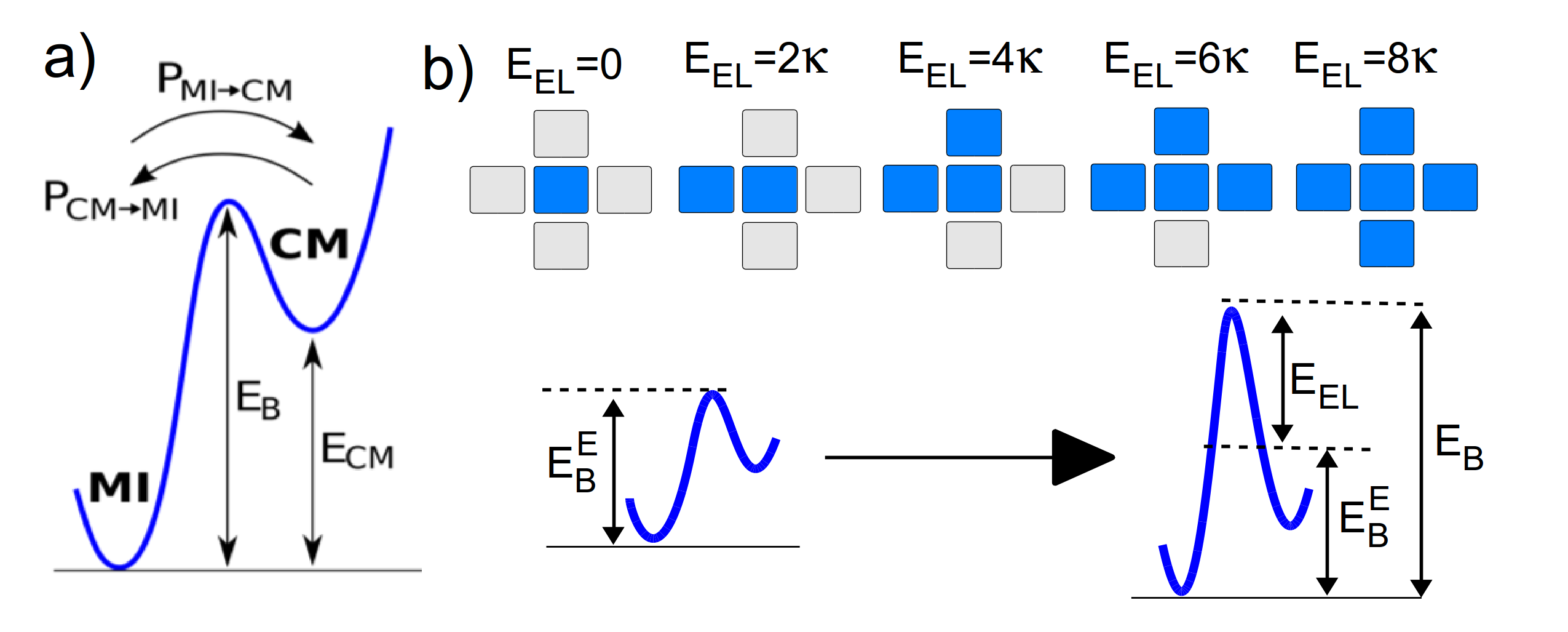}
  \caption{ a) Energy landscape of the model. The Mott-insulator (MI) and the correlated-metal states (CM) are separated by the energy barrier $E_B$. b) The height of the barrier is affected by local structural distortions according to the elastic energy cost of Eq.\ref{Eq_E_el}. We show the different possible configurations for a first neighbor interaction with their respective energy values.}
   
  \label{Fig_energy}
\end{figure}

We start from the model introduced by Stoliar et al. \cite{Stoliar2013} that consists of a 2D resistor-network where 
each element of the network represents a small (nanoscale) region of the physical Mott system \cite{Stoliar2013, Stoliar2014}. 
The cells of the network are assumed to be large enough so that its
electronic state is well defined. Imaging technique experiments with nanoscale spatial resolution have shown that across the 
Mott transition there is a coexistence of metallic and insulating phases with inhomogeneous distributions down to the few 
nanometers  \cite{Basov2007, Dubost2013}.
Each cell site is assumed to be in one of two electronic states: Mott-insulator (MI) or correlated-metal (CM). 
These states are respectively associated with a high and a low resistance values, $R_{MI}$ and $R_{CM}$. 
The resistor network model is schematically shown in Fig.\ref{Fig_intro}.a-b. 
Since the experimental systems are normally in the insulator state, the model assumes that the MI-state is the lowest in energy, 
which is defined as the reference $E_{MI} = 0$. The CM-state is assumed to be 
a { \it metastable } state, with a higher energy $E_{CM}$ , and separated from the MI-state by an energy barrier $E_{B}^E$ of
pure electronic origin (see Fig.\ref{Fig_energy}.a). 
From Dynamical Mean Field Theory calculations of the Hubbard model for Mott systems \cite{Camjayi2014,RMP96},
one may expect that the energy difference between these states may be of the order
of a few tens of $meV$, substantially smaller than the electronic bandwidth .  

In the present numerical study we follow the methodology of Ref.~\cite{Stoliar2013}. 
In analogy to the actual experimental setup, we adopt an electric circuit with a constant {\it load} resistance $R_L$, as shown in
Fig.\ref{Fig_intro}.a. Thus, the resistance of the system $R_S$ and $R_L$ form a voltage divisor circuit.
Denoting $V(t)$ the externally applied voltage protocol, then the voltage on the resistor network 
is given by $V_S(t) = \frac{R_S(t)}{R_S(t) + R_L}V(t)$.  
The external voltage can be set as any arbitrary profile $V(t)$, which upon discretization of the time
becomes $V(t_i)$ at each time-step $t_i$. Initially, all lattice cells are assumed in the MI state.
Then, at every time step $t_i$ the full resistor network circuit is solved, obtaining all the local voltage 
drops $\Delta V_{ij}$ at each cell site ${ij}$, and the total resistance $R_S(t_i)$ between the top and bottom electrodes.
For simplicity the electrodes are assumed perfectly metallic (see Fig.\ref{Fig_intro}.b) \cite{Stoliar2013}.

At each time step, the state of the resistor network is updated following the procedure of Ref.~\cite{Stoliar2013}.
The probability for a given insulator cell to undergo a local EMT, that is from MI to CM at time step $t_i$
is given by $P_{MI\rightarrow CM} = \nu e^{-(E_B^E -eˆ†\Delta V(t_i)/kT)}$ , where the 
constant $\nu$ is an attempt rate, $e$ is the charge of the electron, $ˆ†\Delta V(t_i)$ is the computed local voltage drop at 
the given cell, $k$ is the Boltzmann constant and $T$ the temperature. This is an Arrhenius like law, where the key feature is that
the local electric field increases the probability for the cell undergoing a local EMT. This assumption should (hopefully) be
fully supported by out of equilibrium many body calculations.  
  
A crucial ingredient of the model is the metastability of the CM-state. In fact, if a given cell undergoes a 
MI$\rightarrow$CM transition, then the metastable CM-state may relax back to the MI one.
The transition rate for the  relaxation is also given by an Arrhenius like expression 
$P_{CM \rightarrow MI} = \nu e^{ -(E_B^E-E_{CM})/kT}$, hence the model has a fixed relaxation time $\tau_R$.
We note that the in this relaxation process the field $e\Delta V$ may be omitted since $R_{CM} << R_{MI}$, 
so the voltage drop effect is neglected in the CM sites. 

Thus, in the numerical simulation the 2D resistor network is solved at each time-step $t_i$ 
for a given protocol $V(t_i)$. The local voltage drops $\Delta V_{ij}(t_i)$ are
computed for all cells ${ij}$.  The state of the cells is updated according to the above probabilities. 
This gives a new value of the total (two point) resistance $R_S(t_i)$ of the system, and then the simulation proceeds.
We adopt $\nu = e = kT = 1$. The unit of time in our simulations is the discretized time-step and the 
voltage is in arbitrary units.

\begin{figure}[h]
  \includegraphics[width=0.5\textwidth]{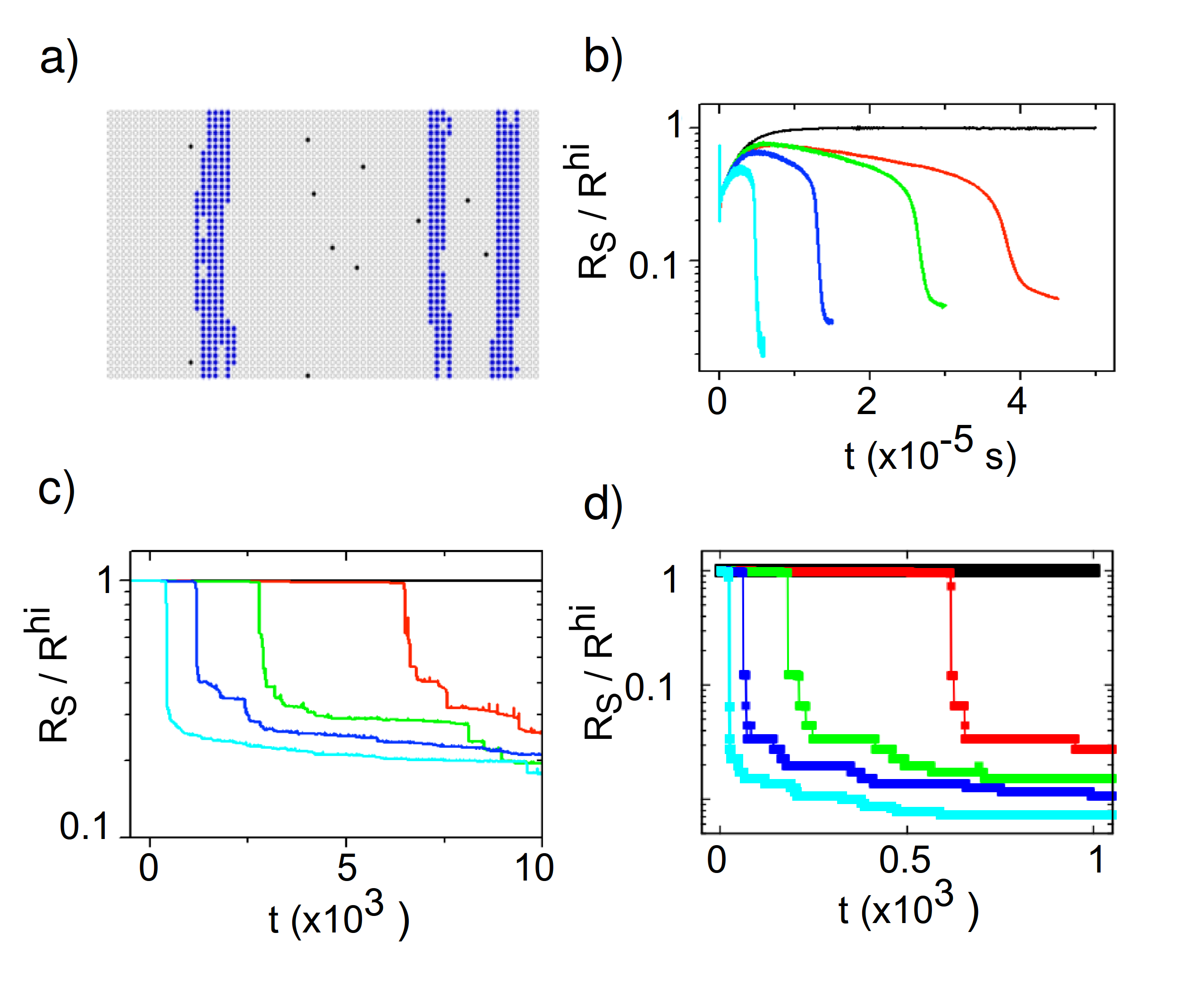}
  \caption{ a) Snapshot of the simulated 2D resistive network after EMT, adapted from ref.\cite{Stoliar2013}. The black and blue dots indicate the sites that are in the CM state, while the gray dots correspond to the sites in the MI state. The dots forming filaments are coloured for easier visualization. The formed filaments exhibit a rather homogeneous thickness as explained in the text. Obs: A different colour-code will be used in the work for the 1D model simulations to avoid confusions. b) Experimental RS curves for different applied voltages in a GaTa$_4$Se$_8$ device. 
  The curves correspond to voltage values of $6$, $40$, $44$, $56$ and $86$ $V$. 
  The length of pulses was limited to $50$, $45$, $30$, $15$ and $6$ $\mu s$. 
  The experiments were performed at a temperature of $T=77K$. 
  The curves show an initial transient increase due capacitive effects. c) Numerical simulations from the original 2D model adapted from ref.\cite{Stoliar2014}. The simulations were performed for a $128$x$40$ network ($N$x$W$) with $E_B^E=20$, $E_{CM} = 10$, $E_{MI} =0$, $R_L=0.1443$, $R_{CM} =
  0.3 R_L$ and $R_{MI} = 16.41 R_L$. The curves correspond to $V$  values of $400$, $600$, $650$, $700$ and $750$.
  d) Numerical Simulations from the 1D filament approximation. The simulations were performed on a $50$x$140$ network  with $R_L=500$, $R_{CM}=200$ ($0.4R_L$), 
  $R_{MI}=200$x$10^3$ ($400R_L$), $E_B^E=11$, $E_{CM}=5$ and  $\kappa=1$. The curves correspond to $V$ values of $300$, $450$, $500$, $550$ and $600$. The same parameters will be used for all simulations in the work, except when specifically indicated. The units of time for simulations are expressed in simulation steps and the voltages are expressed in arbitrary units. The same units are used for all the simulations presented in the work.}
   
  \label{Fig_2D_Comp}
\end{figure}

The previous numerical studies \cite{Stoliar2013, Stoliar2014} showed that the resistive
collapse in the EMT is due to the sudden formation of filamentary structures of CM phase, which connect
the two electrodes. These structures grow, as expected, along the electric field lines. 
Less evident was the observation that while
right after the percolation the filament connecting the two electrodes is quite inhomogeneous, it nevertheless
continues to evolve till it rapidly reaches a rather homogeneous thickness \cite{Stoliar2013}.
This is illustrated in panel (a) of Fig.\ref{Fig_2D_Comp}.
Since the numerical solution of the 2D resistor network is, by far, the most time consuming part of the numerical simulation,
motivated by the previous observations, we explore a simplification of the simulation of the model that would dramatically boost 
the computational time.

This feature suggests that the transverse currents, i.e. perpendicular to the direction
of the applied field, may not play a crucial role, at least after some irrelevant brief transient time.
Therefore we numerically explored the possibility of neglecting the transverse currents altogether in our
simulations.
In Figs.\ref{Fig_2D_Comp}.b-d, we show a comparison of typical experimental data of the 
EMT for various applied external voltages and the respective result of the simulations for the 2D model (panel c) and for the case  where the transverse
currents were neglected (panel d). We see that the later simulations qualitatively capture the behavior, very similarly as 
in the 2D model \cite{Stoliar2013}. In the present case, one should consider the thickness of the 1D filament (one cell) 
as a coarse grained version of the previous 2D one. 

This simplification leads to a dramatic speedup of the simulations. 
In fact, instead of solving a 2D resistor matrix, we simply have a collection of 1D series resistor chains, 
which is trivially solved. This enabled us also to consider significantly larger systems and, more important, 
to simulate for longer times. This turns to be crucial for the systematic
study of relaxation effects, which are relatively much longer than the typical time to dielectric breakdown. In addition, it makes possible a simpler analytical insight, which we will develop in the following sections.

This model has already provided valuable insights and very good qualitative agreement with experiments, such
as the existence of a threshold field and the decrease of the delay time of the resistance collapse with the
increase of the applied voltage \cite{Stoliar2013}.
However, as it stands, if fails to capture the existence of different relaxation regimes (ii) and (iii) as described in the Introduction.
In fact, the model has a single intrinsic relaxation time, which is solely set by the 
energy difference between the electronic energy barrier and
the metastable CM-state, $\Delta E = E_B^E - E_{CM}$. 

Thus, in order to qualitatively capture the observed experimental behavior where strong voltage pulsing may 
realize low resistance states that are long-lived, we shall extend the model by making an additional assumption. 
Specifically, we shall introduce the energy cost associated to the formation of a MI-CM boundary interface. 
We may rationalize this as an elastic energy that originates from the structural strain created at the boundary between the two
distinct electronic phases. In fact, experimentally, using STM techniques, it has been observed  a giant electro-mechanical
effect that was interpreted as a strong self-compression of the lattice structure in the metal state \cite{Dubost2013}.
Therefore, we add to the model the assumption of an elastic deformation energy $E_{EL}$, which is a supplementary
energy cost associated to the transition when two nearest neighbor cells would end in different states. In other
words, it is an energy penalty for creating spatial inhomogeneities.
As we shall see, this assumption creates {\it spatial correlations} in the way that multiple filaments grow. In our previous
model, filaments would grow at random, essentially uncorrelated positions, across the system. However, the addition
of the elastic energy promotes the grow of new filaments neighboring a previous one. Thus promoting the thickening
of filamentary structures.
 
The simplest way to include this in the model is by adding this effect into the definition
of the energy barrier.  

Thus, we may write the energy of the barrier $E_B$ of the $i^{th}$ site of the lattice as

\begin{equation}
	E_{B}(i) =  E_{B}^E + E_{EL}(i)
 \label{Eq_Ecm}
\end{equation}
with the elastic energy cost being,
\begin{equation}
E_{EL}(i) = \kappa \sum\limits_{j}(1 - q'_{i}q_{j})
\label{Eq_E_el}
\end{equation}
where the variable $q_i = \pm 1$ indicates the current state of the $i^{th}$cell (say, MI $= 1$ and CM $= -1$), $q'_i$ is the 
proposed new state for the $i^{th}$ cell (i.e. $q'_i = - q_i$), 
the index $j$ of the sum runs over the number of nearest neighbors $Q$, and $\kappa $ is an elastic constant.

Thus, if initially the $i^{th}$ cell is in the same state as all of its neighbors, then the barrier to overcome is (maximally)
increased from the
electronic value $E_B^E$ by the amount of $E_{EL} = 2 \kappa Q$ corresponding to the maximal mismatch. 
If, on the contrary, the $i^{th}$ cell is initially in a different state with respect to all of its neighbors,
the barrier to overcome is solely the electronic one, as $E_{EL} = 0$ since all cells would end up in the same phase.
For simplicity, we shall consider here that the range of the elastic term is only to the nearest neighbors, 
so in the present 2D lattice $Q = 4$. There is of course freedom to choose longer range interactions. 
We shall come back to this point later on.

As we shall see, the low-resistance states may become long-lived as the filamentary structures grow thick. 
Thus, for the sake of clarity in the presentation of our results, we shall first dedicate the next
section to describe the formation of CM filamentary structures and how their thickness may grow 
under strong applied voltages. Then, in the following section we shall describe how the relaxation
time may be significantly affected by the inclusion of the elastic deformation term and
how it may qualitatively account for the various EMT regimes discussed before. 
We finally also compare our model behavior and
experiments involving the application of trains of short voltage pulses, which bring us qualitatively
closer to the realm of spiking neurons.

\subsection{Filament Formation and Growth} \label{Filament Formation and Growth} 

The EMT proceeds through two distinct stages: the first is the creation of low resistance filamentary 
structures that inter-connect the electrodes; the second is the successive reabsorption 
of the filaments as the system recovers the high resistance state.
The first one occurs under the application of a strong external voltage, 
while the second takes place when the applied voltage
is switched off.
The present section is devoted to describing the first process.

The formation of the filaments under the application of an external voltage in the present model was already qualitative
described in Ref.\cite{Stoliar2013}. The main features were the existence of a threshold field for the formation of filaments
and that the number of filaments (or the fraction of phase change in the system) grows with the magnitude of the applied field
via a process of subsequent multiple filaments formation. The inclusion of the elastic term does not change the 
main basic mechanism, however, there are a few qualitative differences that we shall explore in this section.

In Fig.~\ref{Fig_tD} we show the dependence of the delay time for filament formation $t_D$ with 
the applied voltage on the system ($V_S$), both for experimental data in a GaMo$_4$S$_8$ and for numerical simulations. 
We observe that the model behaves qualitatively similar as the experimental
system, with a variation of $t_D$ spanning many orders of magnitude. 
Both data sets show a linear behavior in a semi-log plot for short times and large $V_S$, and a clear slowing down
of the dynamics as $V_S$ approaches the $V_{th}$.  At the threshold voltage the delay time should diverges, however it is
difficult to access that regime both, experimentally and numerically.

We shall now consider in detail the formation of the filament from a dynamical point of view. This will allow us to
derive approximate expressions for the delay time $t_D$.

The formation of filaments is
a sudden event, thus the state of the system {\it previous} to the EMT breakdown still has the vast majority of cells
in the initial MI state. Thus the system before the rupture is quite homogeneously MI. 
This implies that the role of the elastic term before the rupture is essentially to renormalize the value of the electronic 
barrier height $E_B^E$  to $E_B\approx E_B^E + 2 \kappa Q$  for the MI sites, which upon transition to the CM
state find themselves surrounded by neighboring MI site. In contrast, any site that has already become CM, under relaxation
to the MI does not see any elastic barrier, since all his neighbors are MI.
Then this allow us to understand that the elastic term just produces slight changes in the value 
of the threshold voltage $V_{th}$ and the time  $t_D$ \cite{Stoliar2013}.

One may gain analytic insight in the process of the initial filament formation and threshold behavior
by considering the dynamical behavior
given by the deterministic partial differential equation that follows from the probability rates of our model in a continuum limit.
The production rate of CM-cells is given by the difference between a gain and a loss (leaky) term with their respective
probabilities.

\begin{eqnarray}
\begin{split}
\frac{dn_{CM}}{dt} = [n_{MI} P_{MI\rightarrow CM}] - [n_{CM} P_{CM\rightarrow MI}] \\
                      \\
\approx n_{MI} e^{-(E_B^E + ˆ†E_{EL}^{MI} - V'(n_{CM}))} 
- n_{CM} e^{ -(E_B^E + E_{EL}^{CM} - E_{CM})} \\
                  \label{ncm}
\end{split}
\end{eqnarray}
where $n_{CM}$ and $n_{MI}$ are the time dependent number of cells in the CM and MI state, respectively. $N$ is the
number of cells between the top and bottom electrodes, i.e. the linear longitudinal dimension of the system. 
The elastic energy term in the model depend on the specific local configuration, so for the present analysis we have 
approximated them by the parameters $E_{EL}^{CM}$ and $E_{EL}^{MI}$, which would correspond to average 
elastic energies of a CM and a MI cell, respectively.
The voltage $V'(n_{CM})$ is the voltage drop on the MI cells, $V' \approx V/(N-n_{CM})$. 
Since $R_{MI} >> R_{CM}$, the total applied voltage $V$ essentially falls on the high resistivity 
cells MI, with $n_{MI} = N - n_{CM}$.
Hence, the gain transition rate increases exponentially with increasing $n_{CM}$.

The reader that is not interested in the details of the dynamical analysis of the rate equation that provides approximation
for the delay time $t_D$ (shown in Fig.~\ref{Fig_tD}) may skip to the end of Eq.~\ref{Eq:td2}.

\begin{figure}[h]
   \includegraphics[width=0.45\textwidth]{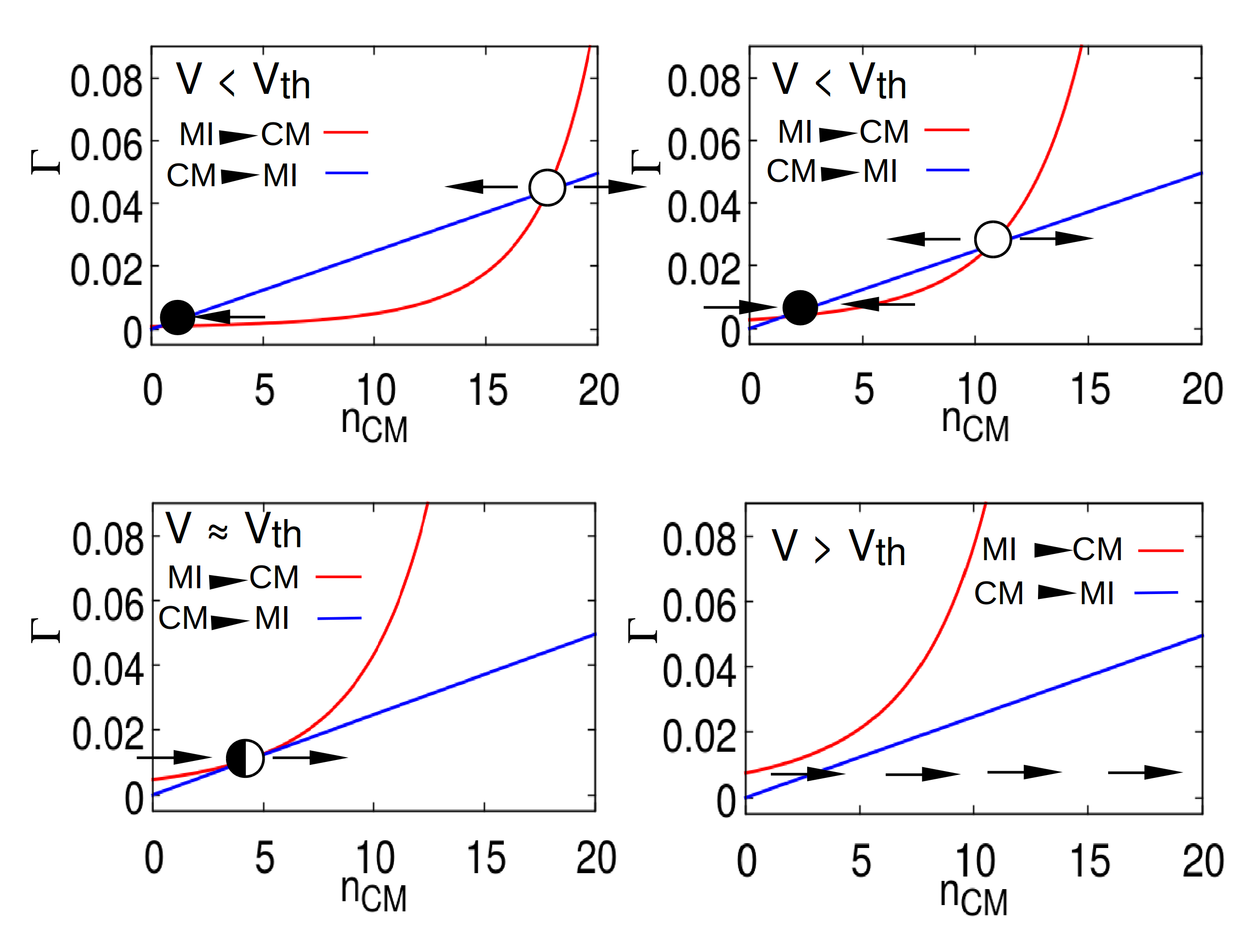}
   \caption{  Dynamical analysis of the model. EMT rates ($\Gamma$) according to Eq.\ref{ncm} as a function of $n_{CM}$. The red curves correspond to the gain rate (first term of Eq.\ref{ncm})  and the blue curves correspond to the lose or ``leaky'' rate (second term of Eq.\ref{ncm}). We show the results for four different applied voltages (increasing from letft-to right and from top-to-bottom panels). For $V< V_{th}$ (top panels) the system 
   exhibits two equilibrium points (where the two curves cross each other). At $V=V_{th}$ (bottom-left) the two equilibrium points ``collide'' at $n^*_{CM}$, while for $V> V_{th}$ (bottom-right) 
   no crossing points are present and no equilibrium can be achieved. The black dots indicate the stable (``attractor'') solution, the empty dots the unstable (``repeller'') solution and the half-filled indicate the collapse of both solutions. In the last panel the 
   arrows indicate the divergent dynamic. This evolution is known as a ``saddle-node bifurcation''.}
   
   \label{Fig_stationary}
 \end{figure}

\begin{figure}
 
   \includegraphics[width=0.4\textwidth]{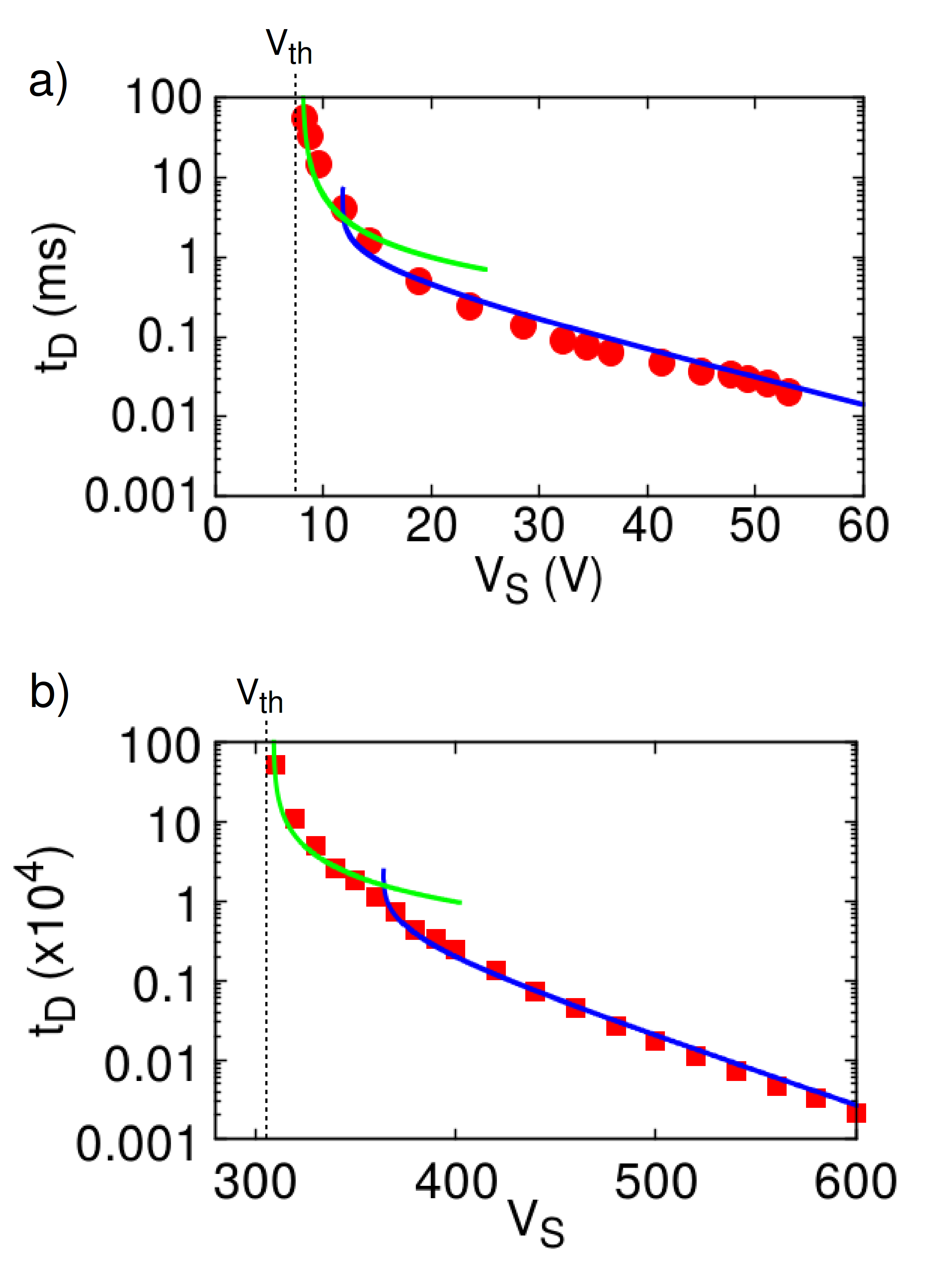}
  \caption[LoF entry]{ Commutation time $t_D$ as a function of the applied 
  voltage. In panel (a) we show experimental results on a GaMo$_4$S$_8$ sample and in panel (b) we show our model simulations. For each case we present two different fits: one for voltages close to $V_{th}$ according to Eq.\ref{Eq:td2} (solid green line) and another fit for the rest of the curve according to Eq.\ref{eq:t1} (solid blue line). The fitting parameters are shown in Table~\ref{tab:param}.
  }
 
  \label{Fig_tD}
\end{figure}

The rate equation has a stationary solution when ${dn_{CM}}/{dt} = 0$. It is given by the balance between the
gain and the loss term. We shall consider the existence of such a solution as a function of the external voltage $V$.
A geometric solution of the problem is shown in Fig.\ref{Fig_stationary} where we plot the gain and loss terms as a function
of the variable $n_{CM}$ at increasing values of $V$ and all other parameter left fixed.
The way to interpret this plot is as follows: at any given value of $n_{CM}$ either the gain (red) or the loss (blue) rate is
bigger. If the gain is bigger, then $n_{CM}$ will evolve to the right, that is, to higher values. If the loss is bigger, it evolves
to the left. The equilibrium of $n_{CM}$ is where the gain and loss transition rates are equal.
We observe in the two top panels of the figure that there are two points where the gain and loss terms cross. 
However, it is not difficult to see that only the lowest one (black dot) corresponds to a stationary {\it stable} equilibrium state,
while the higher one (empty dot) is an {\it unstable} equilibrium point.
Varying the parameter $V$, there is a critical value where the two crossing points ``collide'' at $n^*_{CM}$.
The interpretation of this critical $V$ is the finite threshold value $V_{th}$. For higher
values of $V > V_{th}$, as shown in the last panel, there is no stationary state for $n_{CM}$, which means
that the gain rate is never compensated by the loss rate and there is a runaway evolution of $n_{CM}$ 
(indicated by the arrows). This Instability signals the formation of the first filament.

This process is not an uncommon situation found in dynamical analysis and is termed a saddle-node bifurcation \cite{Strogatz}. For further details, an analysis of the numerical integration of the rate equation is shown in the Supplementary Material.
We shall now derive two approximate expressions for the delay time $t_D$. One is simpler and valid at shorter times and
higher applied $V$ and the other is for longer times and $V$ closer to $V_{th}$.

If the parameters of the model are such that $n^*_{CM} << N$, then one may approximate the rate equation \ref{ncm} by

\begin{equation}
\begin{split}
\frac{dn_{CM}}{dt} \approx
                  N e^{-(E_B^E + ˆ†E_{EL}^{MI} - V/N )} - n_{CM} e^{ -(E_B^E + E_{EL}^{CM} - E_{CM})}
                  \label{ncm2}
\end{split}
\end{equation}
If $V>V_{th}$ then an avalanche event will occur, as described above. Before the avalanche can take place there is a period where the evolution of the system (i.e. of $n_{CM}(t)$) is dominated by the slow dynamics at the proximities of the former stationary point 
where $n_{CM}(t) \approx n^*_{CM}$. 
It is the region where the gain and lose rate curves are approximate parallel (red and blue curves in Fig.~\ref{Fig_stationary}).
Once the evolution of $n_{CM}$ overcomes this regime, it rapidly increases to $N$, i.e. the filament forms with an avalanche.
Thus, we may neglect the duration of the sudden event for the estimation of the delay time for the filamentary formation
and obtain an approximate expression by integration of Eq.~\ref{ncm2} from 0 to $n^*_{CM}$, 
 
\begin{equation}
t_D(V)=\frac{1}{B}ln(\frac{A(V)}{A(V)-n^*_{CM}B})
\label{eq:t1}
\end{equation}
where $A(V)$ is the first term in the right hand side of Eq.\ref{ncm2}, $B$ is the exponential factor in the second term of the
same equation and $n^*_{CM}$ was assumed a constant. 


We used the expression above to produce fits of the experimental and numerical data at short times.
This is shown in blue curves in Fig.~\ref{Fig_tD}, where we observe that the approximate expression 
can reproduce the variation in $t_D$ of more than two orders of magnitude. The fitting parameters are summarized in
the Table \ref{tab:param}. 

At lower $V$, and as we get closer to the threshold, the approximation made before is expected to fail, as mentioned before.
To describe the region close to the threshold we need to more accurately describe Eq.\ref{ncm} expanding around the 
bifurcation point  ($n^*_{CM}$,$V_{th}$). Thus, defining $dn_{CM}/dt \equiv f(n_{CM},V)$, and using that at the bifurcation point $f(n^*_{CM},V_{th})=0$ and $\frac{\partial f}{\partial n_{CM}}(n^*_{CM},V_{th})=0$, we get to the lowest order

\begin{eqnarray}
\begin{split}
\frac{dn_{CM}}{dt}  \approx &\bar{A}(n_{CM}-n^*_{CM})^2+\\ +&\bar{B}(V-V_{th})(n_{CM}-n^*_{CM})+ \\ +&\bar{C}(V-V_{th})
                  \label{Taylor1}
\end{split}
\end{eqnarray}

where


\begin{eqnarray}
\begin{split}
&\bar{A}=\frac{\partial^2 f}{2\partial n_{CM}^2}=\frac{V_{th}^2e^{-(E_B^E+E_{EL}^{MI}-\frac{V_{th}}{N-n^*_{CM}})}}{2(N-n^*_{CM})^3} \\  &\bar{B}=\frac{\partial^2 f}{\partial V \partial n_{CM}}=\frac{V_{th}e^{-(E_B^E+E_{EL}^{MI}-\frac{V_{th}}{N-n^*_{CM}})}}{(N-n^*_{CM})^2} \\ &\bar{C}=\frac{\partial f}{\partial V}=e^{-(E_B^E+E_{EL}^{MI}-\frac{V_{th}}{N-n^*_{CM}})}
\end{split}
\end{eqnarray}

where the derivatives are evaluated at ($n^*_{CM}$,$V_{th}$). Integrating, we obtain for $t_D$

\begin{equation}
t_D= \frac{2atan(\frac{2\bar{A}(n_{CM}-n^*_{CM})+\bar{B}(V-V_{th})}{\sqrt{4\bar{A}\bar{C}(V-V_{th})-\bar{B}^2(V-V_{th})^2}})}{\sqrt{4\bar{A}\bar{C}(V-V_{th})-\bar{B}^2(V-V_{th})^2}} \Bigr|_{n_{CM}=0}^{n^*_{CM}}
\label{Eq:td2}
\end{equation}

The fit according to Eq.\ref{Eq:td2} is shown in solid green lines in Fig.\ref{Fig_tD} for both, experimental and
numerical data. 
We observe that the approximation captures the behavior of $t_D$ near the thresholds. 
The fitting parameters are summarized in the Table \ref{tab:param}. 







We now turn to describe one of the new features introduced by the elastic term in our model, namely
the thickening of the filament with time.
The width of initial filament is is just one cell. Thus, after it forms the resistance of the system $R_S$ drops from a high value
$R_S^{hi} \approx R_{MI}N/W$ to a lower value $R_S^{lo} \approx [R_{S}^{hi}$//$R_{CM}N$] , where $W$ is the width 
of the resistor network and $N$ is the length (i.e. distance between electrodes) as before.
The decrease of $R_S$ provokes a reduction of the voltage applied on the network according 
to $V_S = \frac{R_S}{R_S + R_L}V$,
where $R_L$ is the load resistance. Hence, the new value of $V_S$ may be larger or smaller than $V_{th}$.
In the former case the system will undergo a second filamentary formation, again according to Eq.\ref{eq:t1} but for
a new (lower) value of $V_S$. After the formation of the second filament, $V_S$ will drop further, and the process
of filament formation will continue until $V_S \approx V_{th}$. This feature was already present in the previous model
without elastic term. Thus, in that case, the position where the filaments form was essentially random and
did not have any spatial correlation.

\begin{table}[h!]
\begin{center}
\begin{tabular}{|l|p{1.5cm}|p{1.5cm}|p{1.5cm}|p{1.5cm}|} 
\hline
&\multicolumn{2}{|c|}{Experiments} & \multicolumn{2}{|c|}{Simulations}\\
\hline
\rule[-1ex]{0pt}{3.5ex} &  Region $V\approx V_{th}$ & High Voltage Region &  Region $V\approx V_{th}$ & High Voltage Region \\
\hline 
\rule[-1ex]{0pt}{3.5ex}  $E_B^E$ & 110$meV$ & 100$meV$ & 70$meV$ & 70$meV$  \\
\hline
\rule[-1ex]{0pt}{3.5ex} $E_{CM}$ & 15$meV$ & 15$meV$ & 31.85$meV$ & 31.85$meV$  \\
\hline
\rule[-1ex]{0pt}{3.5ex}  $E_{MI}$ & 0 & 0 & 0 & 0 \\
\hline
\rule[-1ex]{0pt}{3.5ex}  $E_{EL}^{MI}$ & 10$meV$ & 10$meV$ & 49.7$meV$ & 47.8$meV$   \\
\hline
\rule[-1ex]{0pt}{3.5ex}  $E_{EL}^{CM}$ & 0 & 0 & 12.7$meV$ & 12.7$meV$ \\
\hline
\rule[-1ex]{0pt}{3.5ex}  $\frac{n^*_{CM}}{N}$ & 0.08 & 0.05 & 0.07 & 0.04 \\
\hline
\rule[-1ex]{0pt}{3.5ex}  $N$ & 2000 & 2000 & 50 & 50 \\
\hline
\rule[-1ex]{0pt}{3.5ex}  $T$ & 74K & 74$K$ & 74$K$ & 74$K$\\
\hline 
\rule[-1ex]{0pt}{3.5ex}  $V_{th}$ & 8.0 $V$ & - & 1.97$V$ & -\\
\hline 
\rule[-1ex]{0pt}{3.5ex}  $\nu^{-1}$ & $1\mu s$ & $1\mu s$ & 1 & 1\\
\hline
\end{tabular}

\caption{Parameters used for the fits of Fig.\ref{Fig_tD}. For the fit of the simulation data we used as free parameters:
$V_{th}$, $E_{EL}^{MI}$, $E_{EL}^{CM}$ and $n^*_{CM}$. The others parameters are the same as in the numerical simulations. The energy and voltage values were converted from arbitrary units by assuming $T=74K$. For the fit of the experimental data, all parameters are free, but we tried to obtained values physically reasonable.
For instance, for the parameter $N$ that corresponds to the numbers of cells, we implicitly assume a physical cell 
size $\Delta x$=$d/N$=$35\mu m$/2000=$17.5nm$ which is of the order of magnitude of 
the domains observed in STM (see Fig.~\ref{Fig_size}). } 
\label{tab:param}
\end{center}
\end{table}

\begin{figure}[h]
  \includegraphics[width=0.5\textwidth]{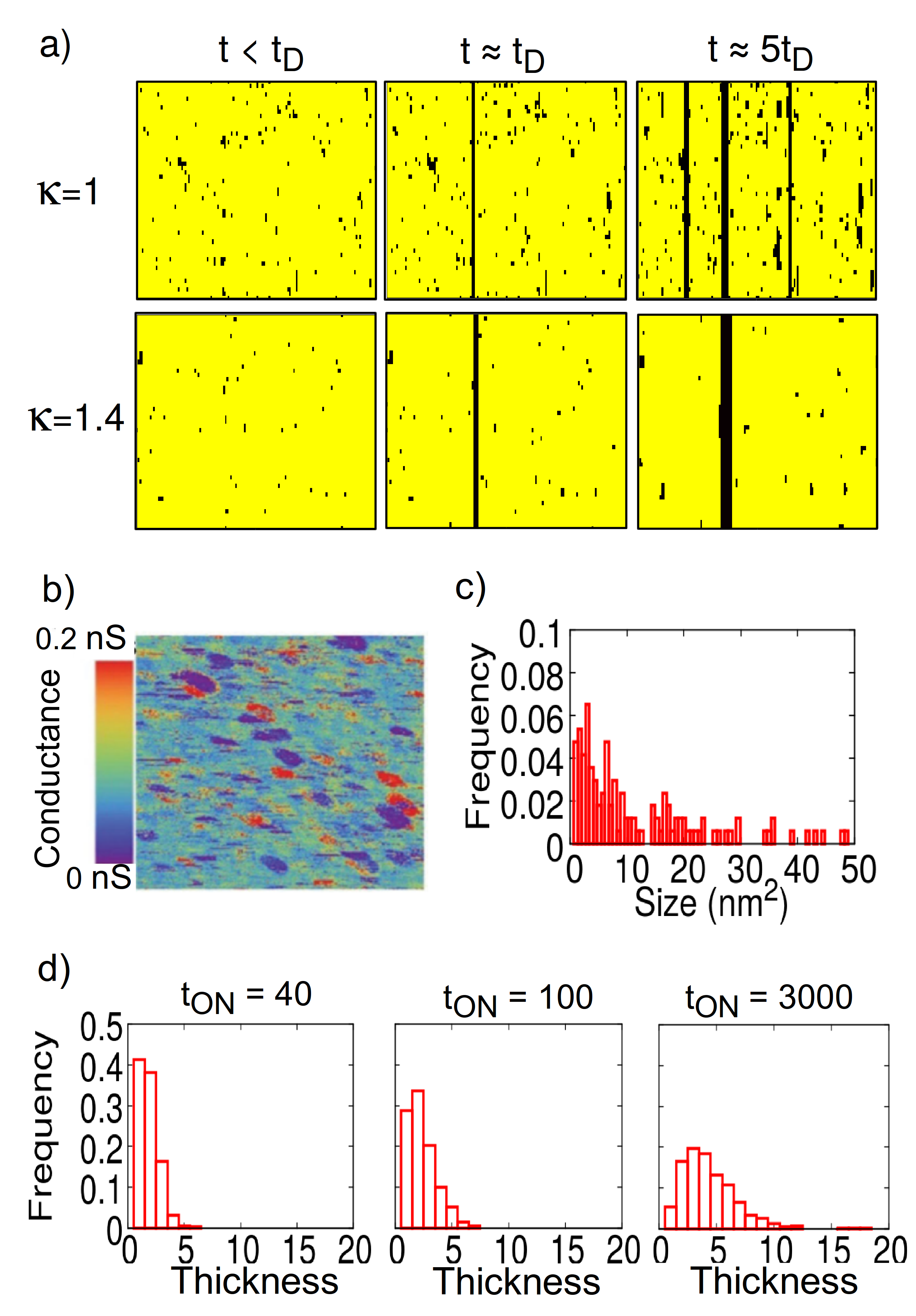}
  \caption{ a) Snapshots of the resistive map during the application of an external voltage ($V=600$). The black dots indicate the CM sites and the yellow the MI sites. We show the results for two different values of $\kappa$. The left panels show the system before the formation of the first filament ($t<t_D$). The central panels correspond to times shortly after the first filament formation. And the right panels
    show the system at times much longer than $t_D$.  
     Higher $\kappa$ favors the thickening process. b) Conductive STM image of a cleaved GaTa$_4$Se$_8$ surface after the EMT 
  (from Ref.\cite{Janod2015}). c) Distribution of cross sectional areas of conductive structures  obtained from (b). d) Distribution of filamentary thickness from simulations. We show the distribution for three different pulse lengths ($t_{ON}$). An applied voltage of $V=600$ was used for the three cases. The data corresponds to an 
  average over 100 realizations.
  }
   
  \label{Fig_size}
\end{figure}

In contrast, in the present model the elastic energy term favors the growth of adjacent filaments, or, in other words, it favors the thickening
of the filaments. This is easy to see, since an isolated filament requires the phase change of cells into the CM-state, while
being surrounded by the other MI phase. That implies a maximal local elastic cost (see above). 
However, if a filament is already in place,
the phase change of a cell adjacent to it is less penalized, hence has a higher probability to occur.
This feature is seen in the panels of Fig.\ref{Fig_size}.a, where we show snapshots of the system before the formation of the
first filament ($t<t_D$), just after its formation ($t \approx t_D$) and at a later stage ($t \approx 5 t_D$). We see that both process have
occurred, more filaments are present but also the filaments grew thicker. Evidently, the higher the value of the elastic
constant $\kappa$ and/or the longer the range of the elastic interaction, i.e. $Q$, the more favorable would be for the 
thickening process with respect to new individual filament formation. Results for two different $\kappa's$ are shown. In addition, the increase of $\kappa$ may enhance the formation of CM-clusters before the generation of filaments, which, in turns, may affect the value of $t_D$. As seen in Fig.\ref{Fig_tD}, this last effect plays little role for $\kappa=1$.

This feature may allow us to interpret the experimental results reported by scanning tunneling microscopy (STM ) on a
cleaved surface of a GaTa$_4$Se$_8$ compound \cite{Janod2015} after the EMT. The image data in Fig.\ref{Fig_size}.b
show the variations in the conductance map due, presumably to the cross sections of the conductive filamentary structures 
that were created by the EMT (in red). The distribution of areas of these conductive cross sections, which we associate
to the filament thickness in our model, are quantified in the histogram in Fig.\ref{Fig_size}.c. 
We can compare these data to our model simulation results. In the three panels of Fig.\ref{Fig_size}.d we show the evolution of
the histogram of filamentary thickness distribution during the application of a continued pulse voltage. We observe that,
consistent with our previous discussion, the histogram distribution slowly drifts to higher thickness and is in qualitative
agreement with the experimental data. We notice here that when typical thicknesses are greater than about $0.1W$ ($W$ being the width of the sample) the nucleation of separated filaments has shown to be a relevant effect (i.e. the creation of big filaments via the percolation of smaller ones). In the STM image we observe cross sections with areas up to about $0.01A$ (being $A\approx500$x$500$ $nm^2$ the total area), suggesting that is effect is not relevant for the experiments analyzed in this work 
\footnote{For a better comparison with simulations, in the histogram of Fig.\ref{Fig_size}.c we consider cross sections up to a few tens of $nm^2$. The full distribution is shown in Supp. Mat.}. 
Thus, we will restrain our study to filaments thicknesses below $0.1W$.

\begin{figure}[h]
  \includegraphics[width=0.5\textwidth]{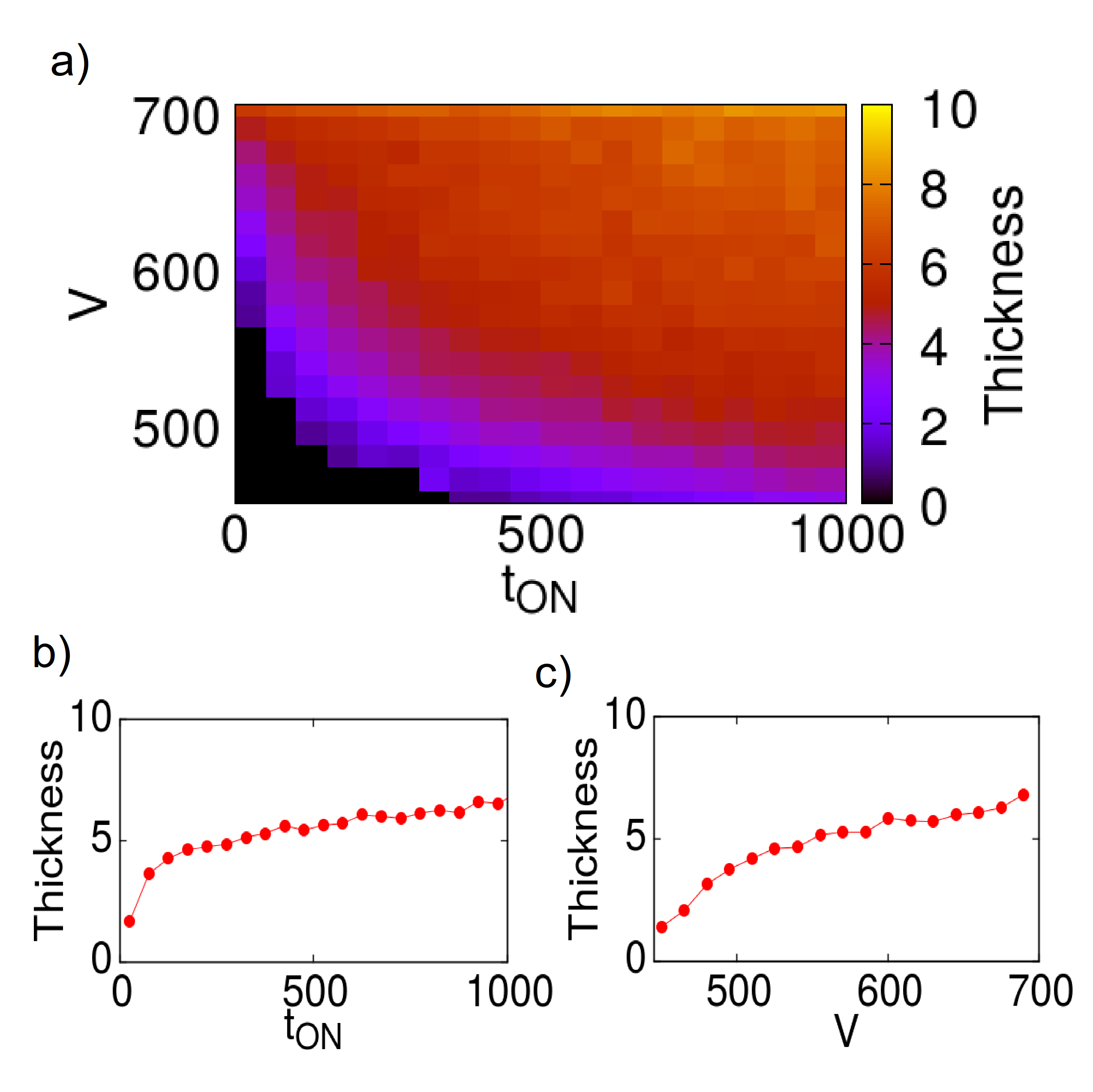}
   \caption{Filament Growth. a) Colormap of the filamentary thickness 
   as a function of applied voltage and pulse length ($t_{ON}$). 
  The results are obtained as the average over $100$ realizations. 
  The thickness is measured considering the thickest filament on each realization. 
  The parameters used for the simulations are the same as in Fig.\ref{Fig_2D_Comp}. 
  b-c) Vertical and horizontal cross sections of panel (a): maximum filamentary 
  thickness as a function of pulse length (panel (b)) and applied voltage (panel (c)). 
  The later was taken after a fixed $t_{ON}=500$ and the former for fixed $V=600$.
      }
        \label{Fig_size_2}
    \end{figure}

The systematic thickness grow of filaments in our model is further characterized in Fig.\ref{Fig_size_2}.
We show a color map (panel a) of the largest thickness of the formed filament as a function of 
the duration of the applied pulse voltage ($t_{ON}$) and its magnitude ($V$). 
The two panels b and c respectively show cuts of the color map at fixed $V$ and fixed $t_{ON}$ 
The results are obtained as the mean over 100 realizations.
The parameters used for the simulations are the same as in Fig.\ref{Fig_2D_Comp}.

\subsection{Filament reabsorption: retentivity in the recovery phase} 

We now turn to a central part of this study, which is the relaxation of the low resistance state back into the Mott insulator
state after the applied voltage pulse is terminated. The proper description of this process is crucial for the understanding
of the recovery phase in neuromorphic applications of the EMT for neuron analogue electronic 
devices \cite{Stoliar2017}.

The previous modeling of the EMT, which did not include the elastic term, predicted that the conductive low resistance
state always returns back to the high resistance state with an essential constant relaxation-time rate. The
cells in the CM metastable state relax back to the MI with a single characteristic time and they do that randomly.
So the filaments disintegrate randomly as well. We shall now study how the addition of the 
electro-elastic effect in the model modifies this feature.

In a low resistance state there are filamentary structures which are composed of a relatively high number of cells 
that switched into the metastable CM-state. These structures are induced by an external applied voltage.
After the voltage application is terminated, the CM cells start to relax back to the high resistivity MI-state. 
The relaxation time for the {\it total} number of cells to return to the MI-state is long. 
However, this is not the physically relevant time for the experiments. 
In fact, the experimental observation of the relaxation of the system is through the behavior of the
recovery of the resistance $R(t)$. So the important quantity to monitor in our model is when the filamentary
structure of CM-cells breaks up or loses the percolation.

In our previous model, the cells forming the filament relaxed in a random fashion. In contrast, in the
present case where we added the electro-elastic term in our model, we shall see that the filaments relax differently.
Rather inversely to the case of filamentary growth by thickening, the dissolution of filaments occurs by thinning.
In fact, as we shall see, the cells on the border of the filament have higher probability to relax to the MI-state
compared to those in the interior of the filament.

In order to perform a systematic study of this process, we shall start from ``artificially'' well defined initial states consisting of 
homogeneous filaments of cells in the CM-state that we let relax back to the insulator. We shall consider initial
filaments of different thickness and then record the time it takes for them to break up. 
Since in our model this process is subject to 
statistical fluctuations we average our results over a large number of realizations (typically 100).
We shall denote this time $\tau_R$ as it indicates the time that a filament keeps its high conductive 
state till it relax to the insulator one. This quantity may also be denoted {\it retentivity}.

\begin{figure}[h]
 \includegraphics[width=0.5\textwidth]{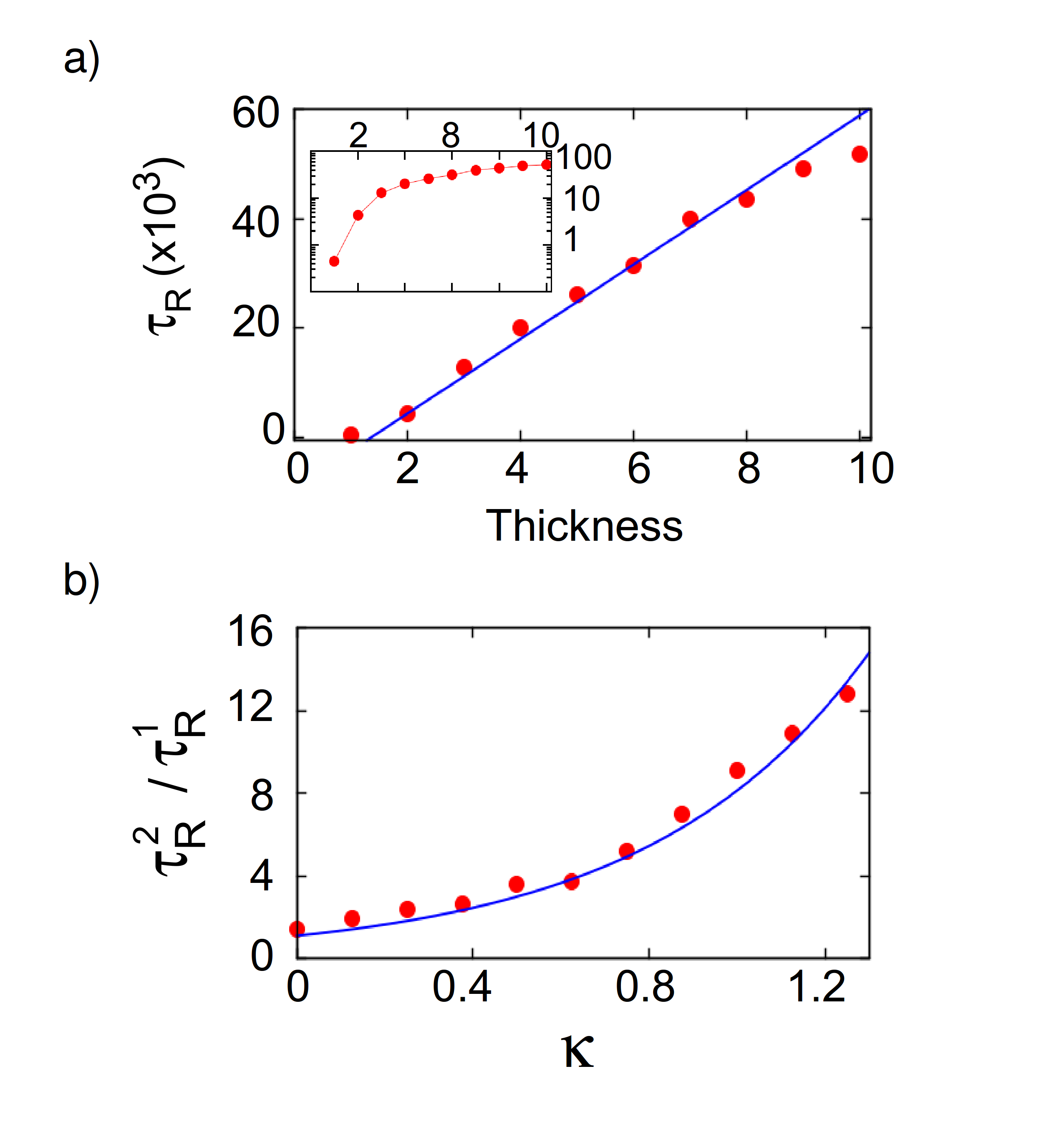}
  \caption{Top panel: Retentivity time as a function of filamentary thickness. 
  The results are the average over 100 realizations. A fit according to 
  Eq.\ref{eq:tret_cluster} (solid blue line) is shown (see Sup. Mat for 
  details on the fit). Inset: Semi-log scale version of main panel. A difference of about one order of magnitude can be seen between the first two points. Bottom panel: Retentivity ratio between filaments of 
  thickness $d=2$ and $d=1$ (red squares) together with a fit according 
  to Eq.\ref{eq:tr_12}: $f(\kappa)=Aexp(2\kappa)$, with $A$ the only fitting parameter. The fit corresponds to $A=1.15$.   }
   
  \label{Figtret}
\end{figure}

In Fig.\ref{Figtret} (left panel) we show results of our simulation studies for the dependence of $\tau_R$ 
with the initial filament thickness $d$, between $d=1$ and $d=10$. 
We observe that there are three different regimes. An initial exponential increase, which then becomes linear
and finally saturates for higher values of $d$.
Overall, $\tau_R$ increased by two orders of magnitude for filament thickness of just a few units.
The fast initial increase, of a full order of magnitude when comparing $d=1$ to $2$, follows from the
presence of the elastic term. This renders much less probable the relaxation of cells in any one of the
two adjacent filaments with respect to a single one. In the former case, any cell of a filament is
surrounded by $3$ other cells of the same type, while in the later only by $2$ (out of $4$). This effect
enters into an exponential in the probability (see Model section above).
As $d > 2$ one may distinguish between central filaments or border filament. Cells belonging to central
filaments have all their $4$ neighbors in the same state. Hence the eventual elastic mismatch maximally 
penalize the relaxation. In contrast, cells in the external boundary of the filament only have $3$ neighbors
in the same sate. Hence boundary cells will relax faster. An internal cell would only get a good probability to
relax by becoming part of the boundary, i.e. after outermost neighboring cells relax. This feature leads to a
continuous ``thinning'' effect of the filaments, and it progresses at an approximate constant rate, which results
in the observed linear dependence with $d$. At higher values of $d$ a saturation effect becomes evident, as
the thinning time starts to compete with the probability for relaxation of internal filament cells (i.e. with
those inside an homogeneous CM phase). 

We may easily estimate the relaxation time of a unitary filament from the loss term of the dynamical
evolution of the cells of expression (\ref{ncm}),
 
\begin{equation}
\frac{\partial n_{CM}}{\partial t} =- n_{CM} exp[-(E_{B}^E + E_{EL}^F - E_{CM})]
\label{eq:ncm}
\end{equation}
being $E_{EL}^F$ the elastic part of the energy barrier of the $i^{th}$-cell in the filament, which is assumed constant for all the cells in the same unitary filament.
The rupture condition for a unitary filament is given by the relaxation of a single cell (since we neglect transverse
currents), i.e. $\Delta n_{CM}=1$. 
A conductive individual filament has initially $n_{CM}=N$. Therefore,
 \begin{equation}
 \Delta n_{CM}={\Delta t} N exp[-(E_{B}^E + E_{EL}^F - E_{CM})] =1
 \label{eq:rupt}
 \end{equation}
and we simply get:

 \begin{equation}
  \tau_R^{F}=\frac{exp[(E_{B}^E + E_{EL}^F - E_{CM})]}{N}
  \label{eq:tr_F}
  \end{equation}
where $\tau_R^F$ denotes the time for rupture of a unitary filament.
From our previous qualitative discussion we get that $E_{EL}^F=4\kappa$ for cells in an isolated
filament of $d=1$. Then $E_{EL}^F=6\kappa$ for cells in the outer boundary of filaments of $d \geq 2$,
and $E_{EL}^F=8\kappa$ for cell in the interior of a thick filament ($d \geq 3$).
The exponential dependence is explicit in Eq.\ref{eq:tr_F}.
 
Then, recalling from our discussion that the filaments relax by progressive thinning, we may simply estimate 
the total time for the full rupture of a filament of thickness $d$ as,
 
  \begin{equation}
 \tau_R \propto d ~ \tau_R^F
 \label{eq:tret_cluster}
 \end{equation}
where $\tau_R^F$ refers here to unitary border filaments.
From the above equations we may focus on the simple case with $d=1$ and $2$, which produce the
largest relative retentivity gain. The ratio for the respective $\tau_R^d$ gives
\begin{equation}
   \frac{\tau_R^2}{\tau_R^1}=exp(2\kappa(Q_2-Q_1)) = exp(2\kappa(3-2))
   \label{eq:tr_12}
   \end{equation}
which corresponds to the blue solid line fit of Fig.\ref{Figtret}.b. Here $Q_d$ correspond to the number of neighbours in the CM state for each thickness $d$. This expression highlights the
exponential dependence on the elastic constant parameter $\kappa$. Evidently, in systems
where the field induced dielectric breakdown is due to a structural change, one may
consider that $\kappa$ is very large and the change cannot be undone except by means of
inducing a new structural change. We may then consider that $\tau_R$ becomes effectively
divergent. Interestingly, this is the situation in non-volatile resistive switching oxides, such as
TiO$_2$, HfO$_2$, NiO, CuO, etc, that form filamentary structures which involve ionic
migration.
One may also contemplate the possibility that in the Mott systems that presently concern us the regime (iii) where the resistive change may be non-volatile (or very long lived) may also involve a volume change of nano-domain size regions of the crystal. Then, the non-volatility may be due to pinning by strain defects induced by the strong electric field \cite{Janod2015}.

On the other hand, it is intuitive that the retentivity time may be significantly augmented by increasing the range of the elastic term.
One may consider, for instance, next nearest neighbors or any longer finite range. Indeed, elastic strain
involves several lattice units in real materials \cite{Wilkinson2006307} . 
At the level of our model, we find this effect is so severe that, even upon increasing the range of the elastic term 
to the next nearest neighbors, our simulations become prohibitively slow. The mathematical analysis of the behavior 
and dependence of model parameters can be carried on along similar lines as we did above (see Sup. Mat.).
Thus the origin of the long (and possibly infinite) retention times in regime (iii) may be due to long range
strain induced by the filamentary growth.

\subsection{Resistance relaxation with an applied voltage bias}

We shall now discuss an interesting effect due to  the introduction of the elastic energy barrier term in our model
related to the electric stabilization of filaments. In other words, electrically prevent the filament break-up and reabsorption.
In previous sections we have seen that a minimum threshold voltage $V_{th}$ is necessary for filament formation. 
In this section we consider a different issue, namely, what is the minimal low-voltage that needs to be
applied to sustain an already fully formed filament. The interest of this question is that it may be experimentally 
tested as further validation of the present model.

\begin{figure}[h]
  \includegraphics[width=0.5\textwidth]{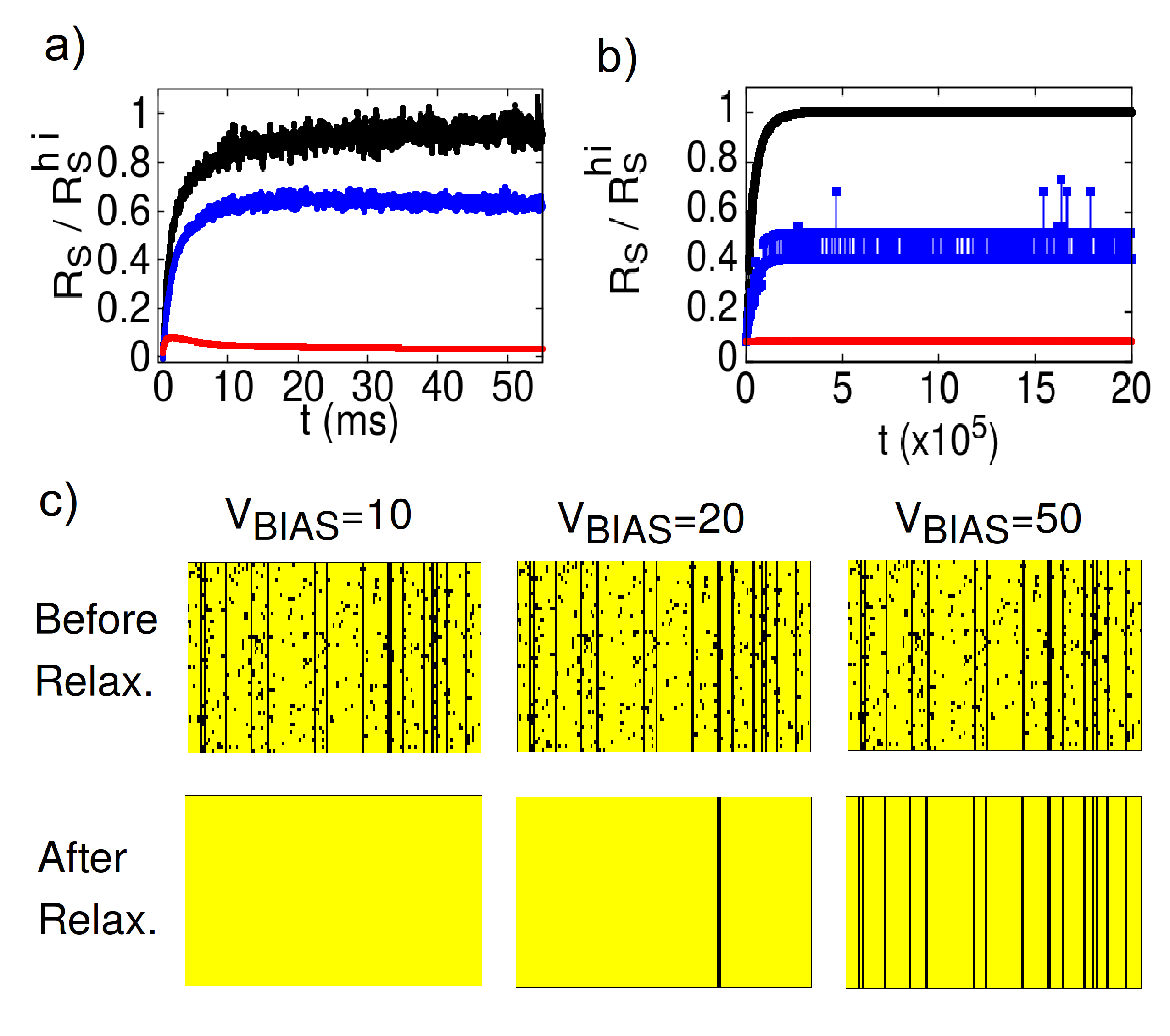}
  \caption{Relaxation of the sample resistance after EMT under different bias voltages. a)Experimental results on a GaMo$_4$S$_8$ sample for $V_{bias}$=2.4, 4.8 and 7.1$V$ (top to bottom). b) Simulations results for $V_{bias}$=40, 50, and 70 (top to bottom). The time is measured from the beginning of the relaxation. For the detail description of the protocol see maintext.  c) Snapshots of the resistive network before and after the relaxation process shown in panel (b).The snapshots correspond to the initial and final states of each curve in panel (b). We see that the state before the relaxation is essentially the same for the three curves. The final state, however, depends strongly on the applied $V_{bias}$. For the lowest bias all filaments are re-absorbed. For the intermediate bias only thicker filaments survive. And for the highest bias all filaments survive. The presence of $V_{bias}$ stabilize (prevent relaxation) of the metallic filaments. Thicker (more stable) filaments require a lower external bias.The simulations  correspond to a $50$x$140$ network with $E_B^E=15$, $E_{CM}=5$, $R_L=500$, $\kappa=0.5$, $R_{MI} = 20000$, $R_{CM}=200$, $t_P=500$ and $t_{ON}=120$. For a better visualization the ratio $R_{MI}/R_{CM}$ was reduced with respect to previous sections. The EMT was generated with an external voltage pulse of $V_{P}=750$ in simulations and $V_{P}=37.5V$ in experiments.}
   
  \label{Bias2}
\end{figure}

In Fig.\ref{Bias2} we show our simulation results for a resistive switching protocol where a
small bias voltage ($V_{bias} < V_{th}$) remains applied during the relaxation process. 
Along with our numerical results we also show in the figure actual experimental data 
obtained in GaMo$_4$S$_8$ sample 
under a qualitatively similar voltage protocol. 
Specifically, the experiment was performed as follows: i) at $t=0$ a bias voltage $V_{bias}$ is turned on, 
ii) at $t=t_P$ a high 
voltage pulse $V_P$ is applied for a time $t_{ON}$, iii) at $t_{OFF}=t_P+t_{ON}$ the applied voltage  returns
 to $V_{bias}$ and the evolution of the resistance of the system is observed. 
 The bias voltage remains applied during the entire process.  The times $t_{P}$ and $t_{ON}$ and 
 the total applied voltage during the pulse $V_{bias}+V_{P}=constant>V_{th}$ are kept constant 
 so that the switching voltage is the same in all realizations. In the figure we show the results of this protocol for three different $V_{bias}$.  We observe that both, in experiments and simulations, the asymptotic relaxation value of the resistance depends on the 
 applied $V_{bias}$. In fact, at the highest applied bias (red curve), we see that the resistance does not recover to the initial
 state but remains at a lower value.
 To get a better understanding of this behavior we show in Fig.\ref{Bias2}.c snapshots of the 
 simulated resistive network for each $V_{bias}$ at two different times: a short one, right 
 after the pulse is terminated at $t=t_{OFF}$  and a longer when the asymptotic resistance state is reached 
 \footnote{As a technical side note we notice that we kept 
 the same ``seed'' for the pseudo-random number generator in all three simulations for the sake of a more meaningful comparison.}.
We observe from the short-time snapshots (upper panels) that the $R_S^{hi} \rightarrow R_S^{lo}$ resistive
change is rather independent of the applied (sub-threshold) bias-level, as a very similar multi-filament 
structures are realized in all three cases 
 On the other hand, for the asymptotic state the system shows qualitatively different states.
 In the lowest bias case, the whole filamentary structure is reabsorbed. In the intermediate bias
 case only a single filament persists, which was the thickest. 
 While in the higher bias case, all filaments formed during the pulse remain at asymptotic times. 
 We also note that all isolated metallic regions (clusters) do relax to the insulator
 state in all three cases.The fluctuations observed for the intermediate voltage in panel (b) (blue curve) are due to rapid rupture and reconnection of small sections of a filament (i.e. a transitory decay to the insulating state). This effect may
 actually happen in real materials, however in our simulations it is
 likely overemphasized due to the finite size.

\subsection{Complete pulse-voltage protocol}

In this final section we shall consider the full-time behavior of the system, that is the whole process of 
formation and subsequent reabsorption of filaments. We shall adopt two different applied voltage protocols:
a single pulse and a train of pulses. The former case is simplest to analyze as is closely related to the
discussions of previous sections. In contrast, the interest of applying train of pulses \cite{Stoliar2017}
is that it bring us closer to the situation of actual neurons, which are excited by electric spikes.

\begin{figure}[h]
  \includegraphics[width=0.5\textwidth]{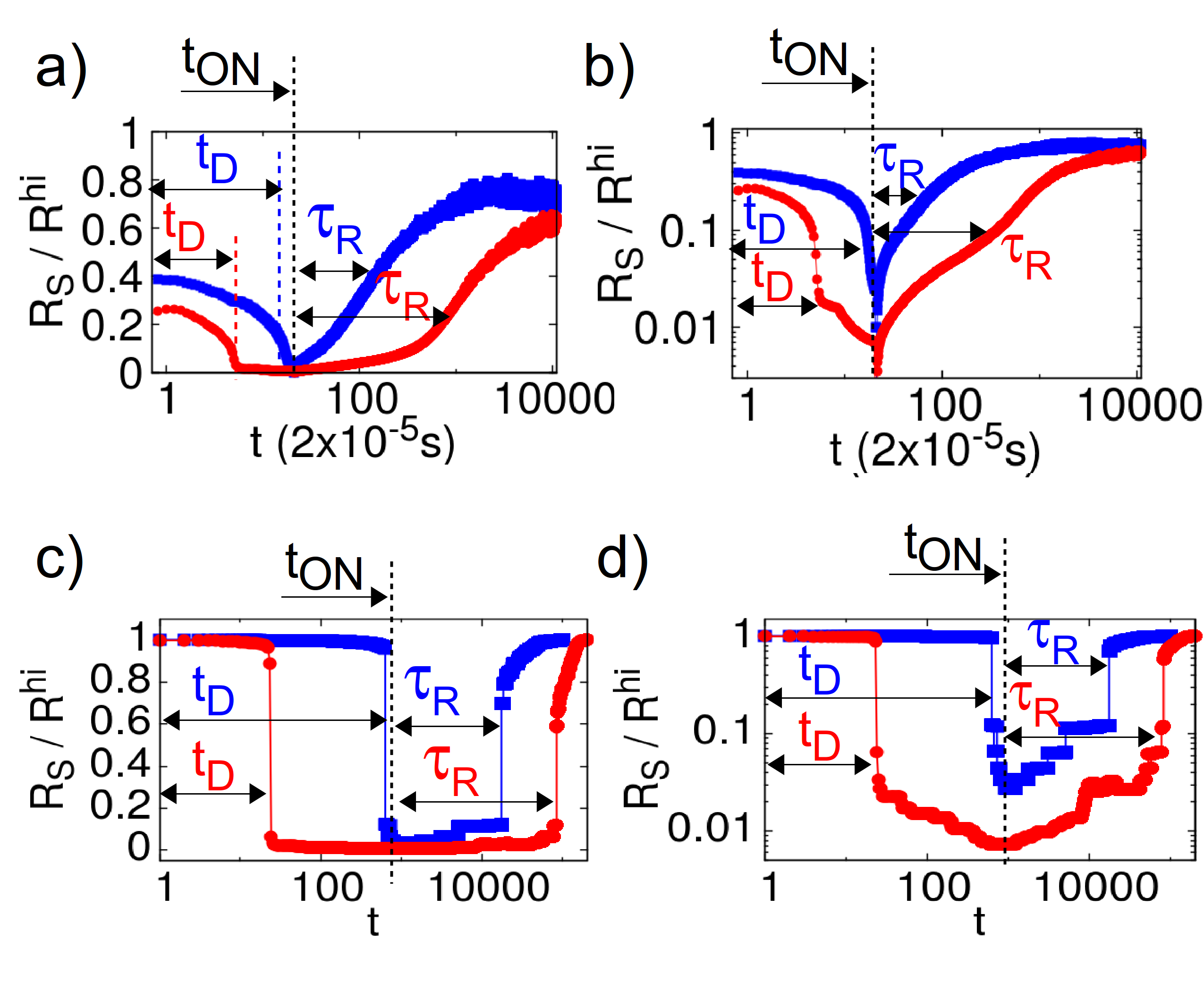}
  \caption{Evolution of the resistance during a complete Switching process. Panels (a-b) show experimental results on a GaMo$_4$S$_8$ sample while panels (c-d) show the results of our simulations in semi-log and log-log scale respectively. In each case an external voltage is appplied for a time $t_{ON}$ after which the system is left to relax freely. The red and blue curves represent the results for two different applied voltages. The experimental curves correspond to $t_{ON}=100\mu s$, $V=34V$ (blue curve) and $V=56V$ (red curve). The simulations correspond to $t_{ON}=1000$, $V=450$ (blue) and $V=600$ (red). The characteristic times $t_{D}$ and $\tau_R$ are indicated for each case. We notice that in the experimental data the initial value of $R_S/R^{hi}$ is smaller than unity due to a capacitive transient introduced by the measurement circuit.}
  \label{Fig_Rvst_1}
\end{figure}

In Fig.\ref{Fig_Rvst_1} we show the full time evolution of the
resistance during the application of a single pulse for both simulations and experiments on a  GaMo$_4$S$_8$ sample (same sample as in Fig.\ref{Fig_tD}) . A single voltage is applied for a fixed duration $t_{ON}$, then $V$ is set 
to zero. We show data for two different applied voltages.  We observe the significant
dependence of the retention time with the applied voltage. We see that a relatively small 
change in the strength of the applied pulse produces a great change of the retention 
time. For example, in experiments a change in the voltage from $34V$ to $56V$ produces a change in the retention time of a full order of magnitude (from approximately $2$x$10^{-3}$ to approximately $2$x$10^{-2}$). Similarly in the simulations, a change in the voltage from $450$ to $600$, produces a change of the retention 
time by a factor of $500\%$ (from approximately $2$x$10^4$ to approximately $10^5$).

\begin{figure}[h]
  \includegraphics[width=0.5\textwidth]{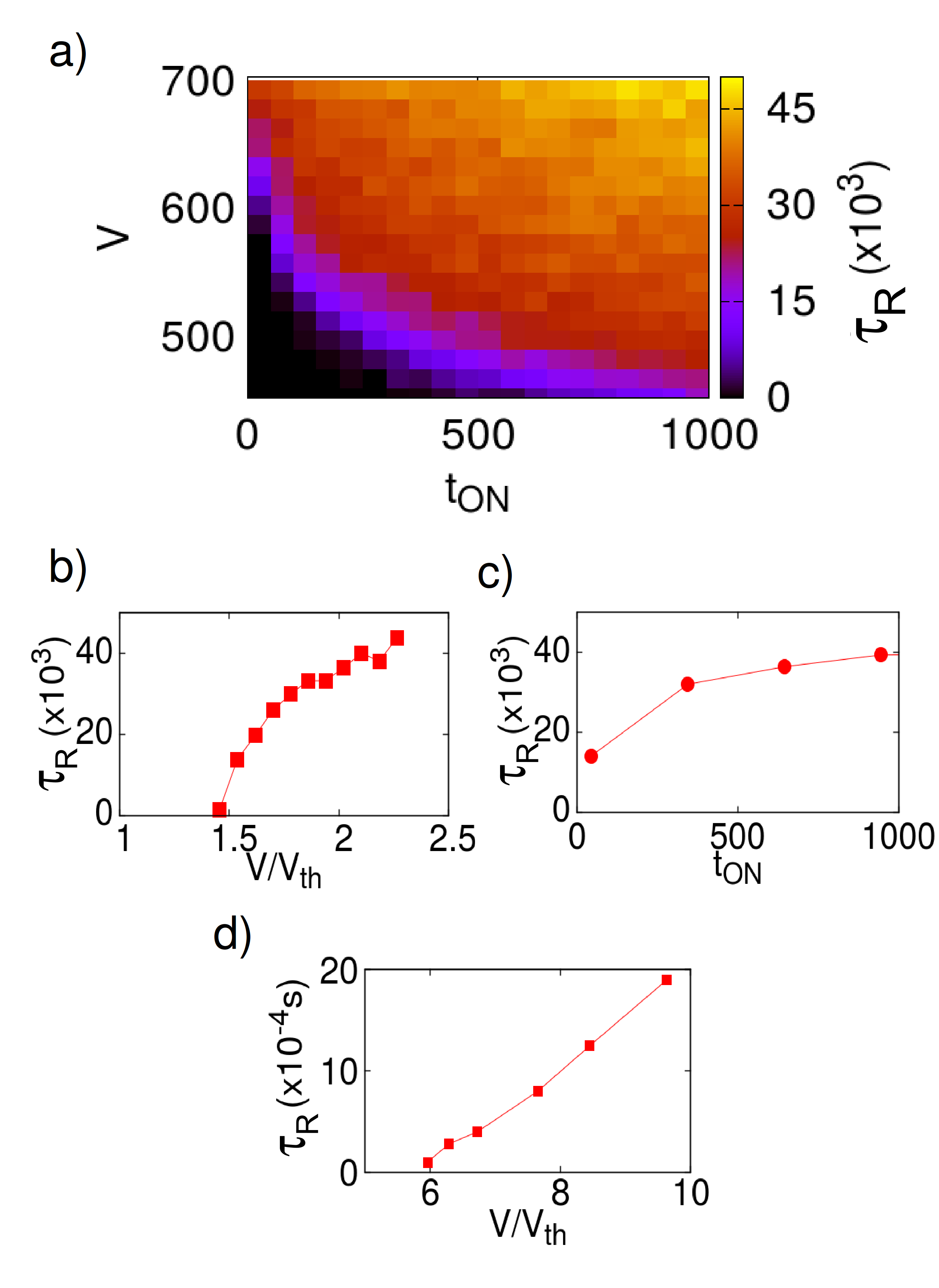}
  \caption{
  a) Colormap of the retentivity as a function of pulse length ($t_{ON}$) and applied voltage. 
  b) Vertical cross section of panel (a): evolution of $\tau_R$ as a function of the applied voltage for fixed $t_{ON}=500$. 
  The voltage is measured in units of the threshold voltage with $V_{th}=309$. c) Horizontal cross section of panel (a): evolution of $\tau_R$ as function of pulse length for fixed $V=600$. The results are obtained as the average over $100$ realizations. d) Experimental results: evolution of $\tau_R$ on a GaMo$_4$S$_8$ sample as a function of the applied voltage for fixed $t_{ON}=100\mu s$.  }
   
  \label{Fig_Rvst}
\end{figure}

These results permit to
rationalize the existence of the regimes (ii) and (iii) that we described in the introductory section before. 
In Fig.\ref{Fig_Rvst} we show our simulation results for the retention time $\tau_R$ as a function of
parameters of a single voltage pulse: intensity $V$ and duration time $t_{ON}$. 
In the color plot of panel (a) we observe that the retention of the filaments may be increased
by either incrementing $V$ or $t_{ON}$. The key point being that any one of these two parameters
contribute to thicken the filaments. The dependence of the threshold with the duration
of the applied pulse can also be clearly observed. This is basically the effect of the delay time that
was discussed in previous sections: At a given applied voltage there is a delay time for the transition, which 
implies that the pulse duration must be above the delay time to actually produce the filament formation. 
The two panels (b) and (c) show two cuts of the color map for fixed $t_{ON}$ and $V$ respectively. All these results correspond to the mean over $100$ realizations. Finally in panel (d) we show the evolution of $\tau_R$ on a GaMo$_4$S$_8$ sample as a function of the applied voltage for fixed $t_{ON}=100\mu s$.

\begin{figure}[h]
  \includegraphics[width=0.5\textwidth]{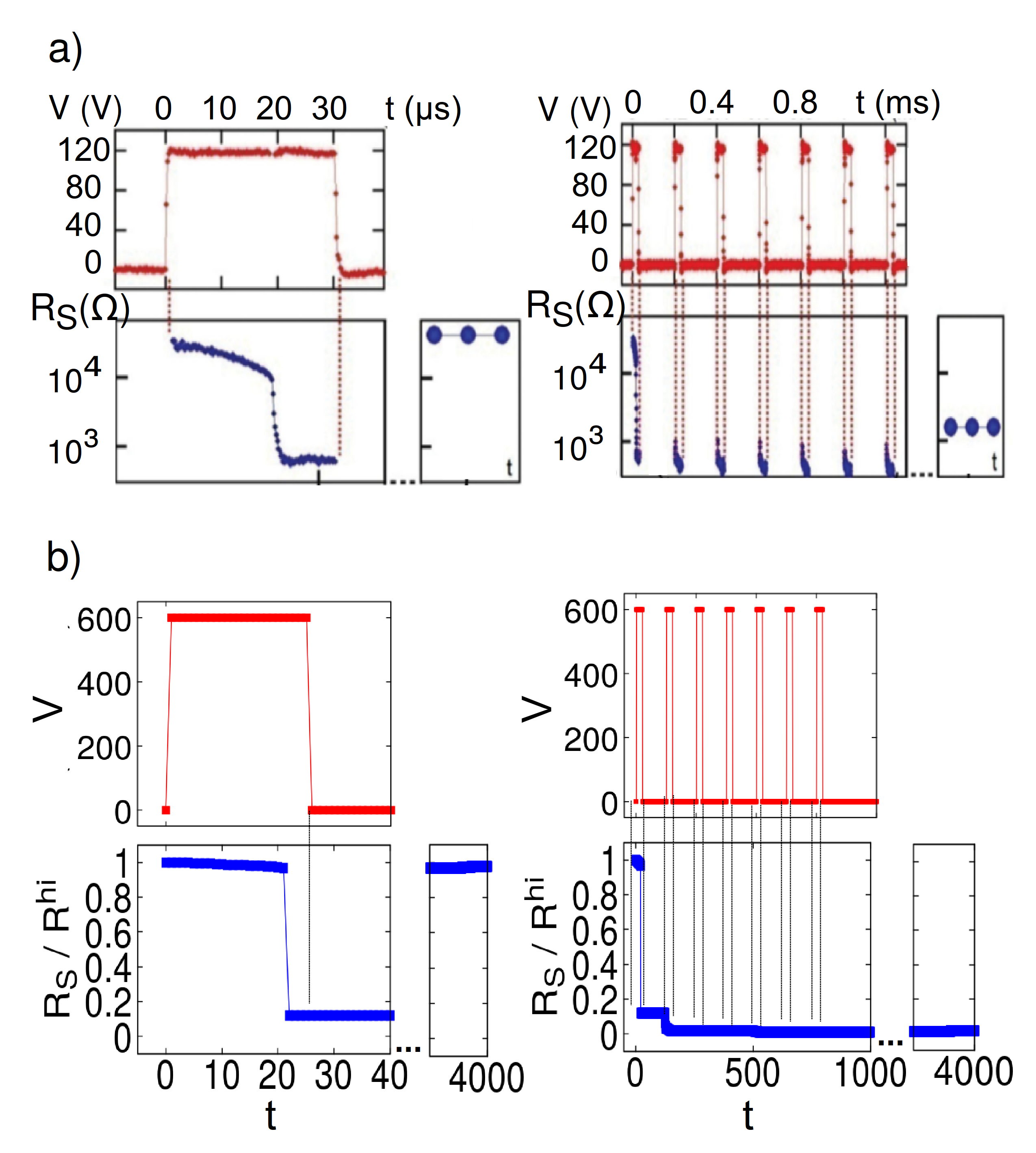}
  \caption{Retentivity manipulation in RS via a pulsed protocol. a) Experimental results in a GaV$_4$S$_8$ crystal for a single pulse of $t_{ON}=30 \mu s$ with $V=120V$ (left), and for a train of $7$ pulses of $t_{ON}=30 \mu s$ with $V=120V$ separated by $200 \mu s$ (right). b) Results from the simulations for a similar pulsed protocol. A single pulse of $t_{ON}=25$ with $V=600$ (left) and a train of $7$ pulses of $t_{ON}=25$ with $V=600$ separated by $100$ simulation steps (right) are shown. }
   
  \label{Fig_pulsed}
\end{figure}

We may finally illustrate the behavior of the model upon the application of a train of voltage pulses.
We show in Fig.\ref{Fig_pulsed} the comparison of our model simulations with actual experimental data of the resistive 
change in a crystal of  GaV$_4$S$_8$ upon application of voltage pulses. 
Two types of pulse protocol are applied: either a single pulse or a 
train of identical pulses with a relatively long separation, which mimics a train of spikes that may 
be arriving to the dendrites of a neuron \cite{Stoliar2017}. In actual situations the arrival of pulses is 
mostly random, though here we focus on regular trains for the sake of simplicity.
In the top panels of the figure we show the experimental data and in the bottom the simulations.
Each pulse of the train is identical to the single one.
The data in red color corresponds to the applied voltage and the data in blue to the respective resistance value $R(t)$.
The small subpanels at the right of the main $R(t)$ panels indicate the asymptotic value of the resistance after the voltage
pulse protocol is terminated. In the experimental data we observe that the single pulse is sufficient
to produce a resistive change, which takes place after about $20\mu sec$. However, this low-resistance state is
volatile, as the corresponding asymptotic value has returned to the original high resistance value.
In contrast, as the data of the right panel show, the application of $7$ identical pulses
was enough to dramatically increase the retention time.
Thus, seven pulses are found to be sufficient to drive the resistance change non-volatile (in the time scale of
the experiment).

In the bottom panels we see that the simulations capture qualitatively well these effects.
From the discussions of previous sections we can understand the behavior.
As shown in the left panels of the figure, a single pulse is enough to induce an EMT with filamentary formation. 
However, that single pulse only produced a thin filament with a short retention time, which got 
rapidly reabsorbed. Thus the system returned back to the initial high resistance state. 
In contrast, the train of multiple pulses produced the gradual thickening of the filament, 
with a consequent much longer retention time and persists beyond $t=4000$ .

This behavior may be considered an initial step towards the modeling of a recovery period of an artificial neuron 
under spiking stimulus. It may be interesting to mention that this variations in the recovery time may find potential useful applications 
in neural networks for tasks such as neural coding \cite{Rieke,Gerstner1997}.

We may note that the inter-pulse time duration ($200 \mu$s in the experiment)
provides a rough indication of the reabsorption time of the filaments. 
The systematic investigation of the retention time with the parameters of the applied voltage pulsing
protocol is an important issue but it certainly involves some technical challenges and 
is outside the scope of the present study. 

\subsection{Conclusions}
In the present study we have extended a model for the EMT to incorporate electro-elastic effects.
This feature lead to spatial correlations in the growth patterns of filamentary structures in the
resistive transition. In particular, it leads to the thickening of the filaments during longer applied
pulses or multiple pulses. The increased thickness of filaments was directly related to a dramatic 
increase in retention time of the low-resistance state. 

We should mention that our model does not predict any sharp or discontinuous transition 
between the experimentally observed volatile regime (ii) and the non-volatile
regime (iii)  discussed in the introduction. Nevertheless, it does indicate that the retention times
may be exponentially increased by means of the applied pulsing protocol.
Specifically the strength, duration and inter-pulse time, which may be tailored to optimize the
growth and thickening of the filaments. The results of the previous section may be considered
as an initial step in that direction. However, it should be realized that there are a large number of 
variables to take into account and the systematic investigation remains an important experimental challenge.

We also briefly refer to the qualitative similarities of our present model study and the behavior
of spiking leaky-integrate-and-fire neurons. We are aware of the relative simplicity of our system with respect to the
biological ones but, nevertheless, this may be taken as a contribution towards future implementation of
neuromorphic electronic systems.
In this regard, it is very exciting to envision that strongly correlated Mott materials may realize
unexpected bio-inspired functionality.

\bibliographystyle{apsrev4-1}
\bibliography{Ref.bib}

\newpage

\onecolumngrid

\section{Supplemental Material}

\section{Dynamical Analysis and filament formation}
\begin{figure}[h!]
  \includegraphics[width=0.6\textwidth]{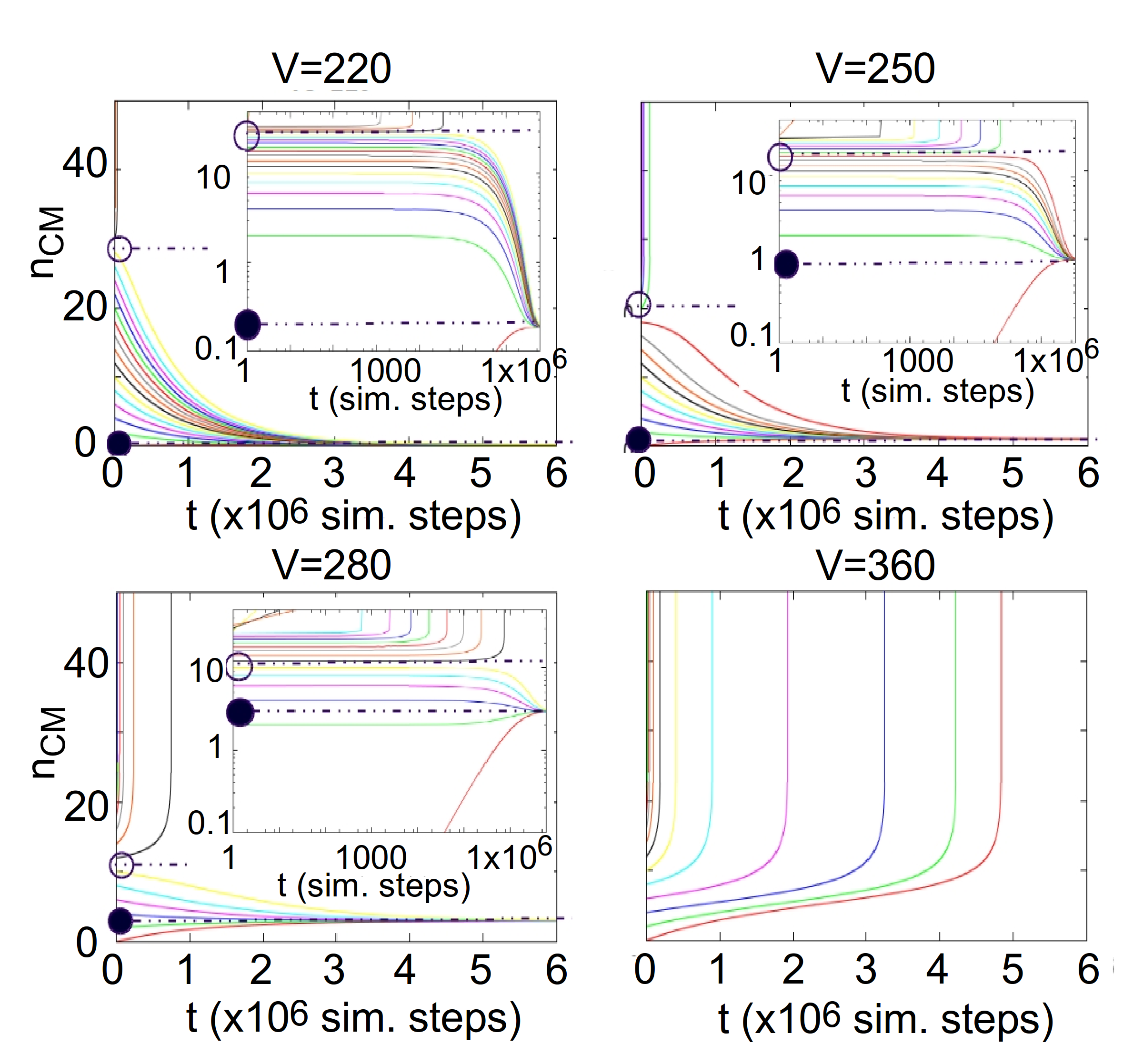}
  \caption{ Dynamical analysis of the model. Numerical integration of Eq.\ref{eq:ncm_sup} via a $4th$ order Runge-Kutta method. It can be seen how the two fixed points (one attractor and one repulsor) gets closer as the voltage is increased until they collapse and disappear.This transition is known as a  ``Saddle-node bifurcation'' \cite{Strogatz}. The insets show the plots in a log scale.}
  
  \label{fig:Bifurc}
\end{figure}

In this section we complement the dynamical analysis of the model made in the maintext. We present a numerical integration of the model equations  that 
By definition, the threshold voltage is the minimum one at which at least one filament is formed. In our description each filament is independent from each other, so the analysis is done studying a single-filament dynamics. To begin we notice that in our model the change in the number of conductive cells at time $t$ is given by the net rate equation of cell metalization (see maintext):

\begin{equation}
\begin{split}
\frac{\partial n_{CM}}{\partial t}=P_{MI\rightarrow CM}-P_{CM\rightarrow MI}= \\
                  n_{MI}e^{-(E_B^E+E_{EL}^{MI} -q\Delta V )/kT}-n_{CM}e^{ -(E_B^E+E_{EL}^{CM} -E_{CM} )/kT}
                  \label{eq:ncm_sup}
\end{split}
\end{equation}

where $n_{CM}$ is the number of metallic sites in the filaments at time $t$ and $n_{MI}$ the number of $MI$ sites (with $n_{CM}=N-n_{MI}$, being N the total number of sites in a filament) and where we assumed that $E_{EL}$ is at this point approximately the same for all sites in a given state (valid for $t<t_{D}$, as explain in maintext). For simplicity we take  and $q=kT=1$ and we will start assuming $E_{EL}=0$. When the right-hand side of Eq.\ref{eq:ncm_sup} becomes zero then the rate of metalization is zero and the system is said to have a ``fixed point'', which represents an equilibrium state of the system. In maintext, geometric solutions of this equation are shown for different applied voltages and the existence of a saddle-node bifurcation is shown \cite{Strogatz}. Here we analyze the dynamics of the system via direct numerical integration.  
In Fig.\ref{fig:Bifurc} the integration of Eq.\ref{eq:ncm_sup} with a 4th order Runge-Kutta method is shown. Here we assume that the voltage applied on the sample is independent of time (valid for $t<t_{D}$), and that the drop of voltage on each cell is given by $\Delta V\approx V/n_{MI}$. Each panel in Fig.\ref{fig:Bifurc} correspond to a different applied voltage. And each curve inside a panel correspond to a different initial condition (different initial $n_{CM}$ ). Here we used that each filament has total length of $N=50sites$ and $n_{CM}=N-n_{MI}$. In this plot we can see the evolution of the two fixed points: the ``attractor'' point (represented with a filled circle) and the ``repulsive'' point (represented by an empty circle). We can see again how the two
points gets closer to each other as the voltage is increased, then collapse (at $V=V_{th}$ ) , and then no fixed points are present. The different initial conditions curves are meant to show the location and character of each fixed point: the system diverges from the ``repulsive'' point and converges toward the ``attractor'' point. The actual ``physical'' case would correspond to an initial $n_{CM}$ close to zero, as the experimental devices begin in an insulating phase. For $Vs=360$ ($V>V_{th}$ ) it can be seen that, before diverging, the system spends a significant amount of time at $n_{CM}\approx5$ (close to where the fixed point was placed right before the bifurcation). This time-delay effect, known as the ``ghost'' of the fixed point, is a regular feature of saddle-node bifurcations and correspond to what we call $t_D$ in the maintext.

\section{Filaments size Distribution in Experiments}

\begin{figure}[h!]
  \includegraphics[width=0.3\textwidth]{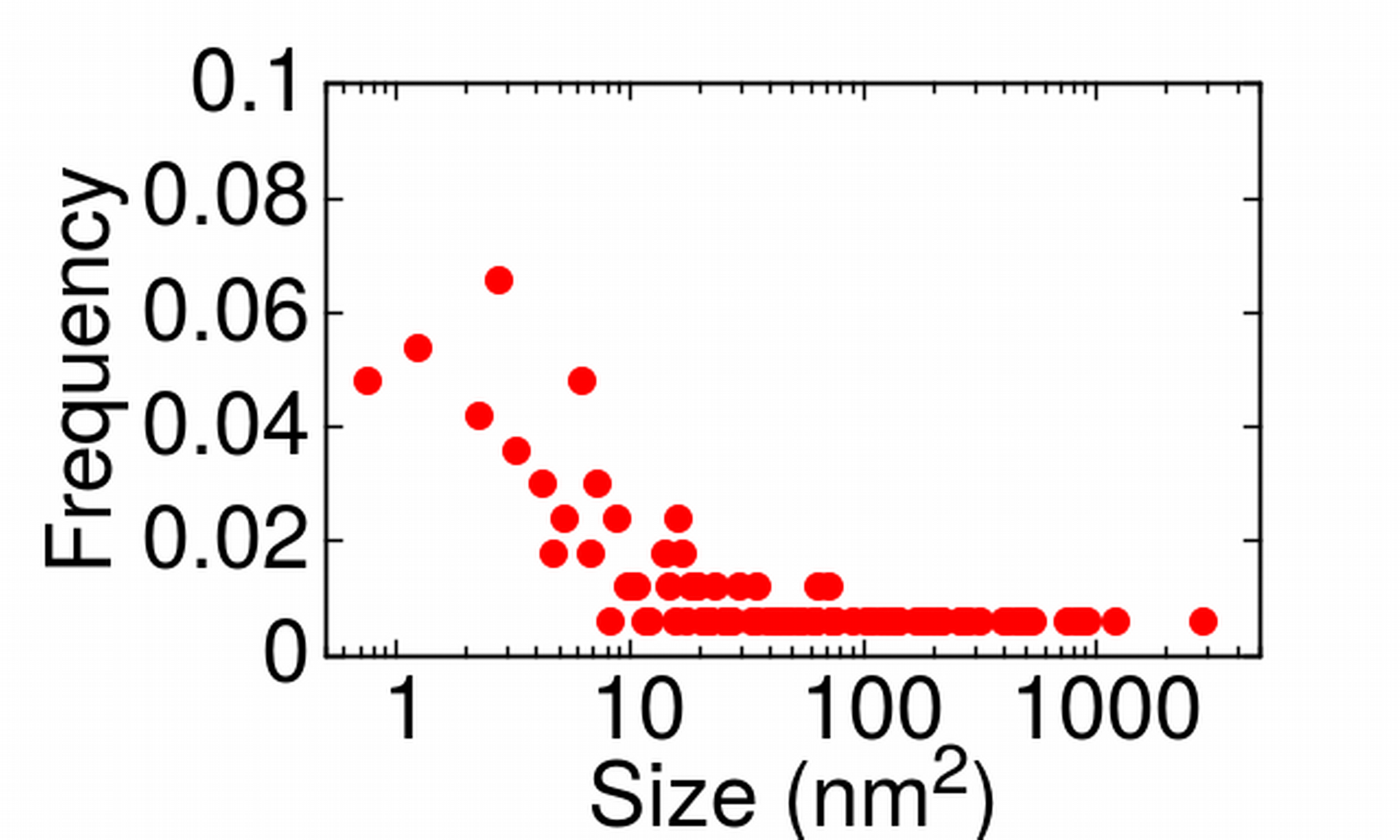}
  \caption{ Distribution of cross sectional areas of conductive structures  obtained from the STM image of a cleaved GaTa$_4$Se$_8$ surface after the EMT (see Fig.7 in maintext). }
  
  \label{fig:STM}
\end{figure}

In Fig.\ref{fig:STM}  we show the full distribution of metallic cross sections obtained from the scanning tunneling microscopy (STM) shown in Fig.7.b of maintext. The image correspond to a cleaved surface of a GaTa$_4$Se$_8$ compound after the EMT. The total area of the sample is of about $500$x$500$ $nm^2$.

\section{Retentivity for clusters with $d>w_B$}

In a filament of thickness greater than range of interaction ($w_B$) we can have two types of cells: the ones in the center and the ones in the ``border''.  Each type will have a different energy  $E_{EL}$. For simplicity we will consider here the case of a n.n. interaction (where $w_B=2$). In this case the cells in the border have three CM neighbors and one MI from where $E_{EL}=6K$  (see maintext). And the ones in the center have 4 CM neighbors from where $E_{EL}=8K$. Using the EMT rate equation we can find the mean lifetime of each layer (central or superficial) $t_{ret-Center}^F$ and $t_{ret-Border}^F$. 
If $t_{ret-Center}^F >> t_{ret-Border}^F$ then the filament will tend to relax from outside-in (starting from the surface towards the bulk) as explained in maintext. After one cell in the surface turns MI then its nearest inner-neighbor becomes superficial, and after it relax the next neighbor becomes superficial and so on. Then the rupture of a filament of thickness $d$ can be thought as the consecutive rupture of $d/2$ superficial cells (considering that the process begins at both boundaries of the filament with equal probability):

\begin{equation}
\tau_R=C\cdot \frac{(d-2)}{2}\cdot \tau_{R-Border}^F + \tau_{R-Border}^F
\label{eq:tret_cluster_sup}
\end{equation}

being $\tau_R$ the retentivity time of the filament and $C$ a factor related with the geometry of the de-percolation path (i.e. shape and size of the outer-in perforation) 
In maintext the evolution of $\tau_R$ as function of the thickness obtained from the simulations is shown, together with fit according to Eq.\ref{eq:tret_cluster_sup}. For the fit we fixed $t_{ret-Border}^F=4400$ (according to the retentivity obtained for $d=2$) and used $C$ as a free parameter, obtaining $C=3.1$.

For a general $w_B$ the description is completely analogous. In the next section we present the results for longer interactions, ranging from 2nd neighbors to 4th neighbors.

\section{Longer range interactions}

\begin{figure}[h]
  \includegraphics[width=1.0\textwidth]{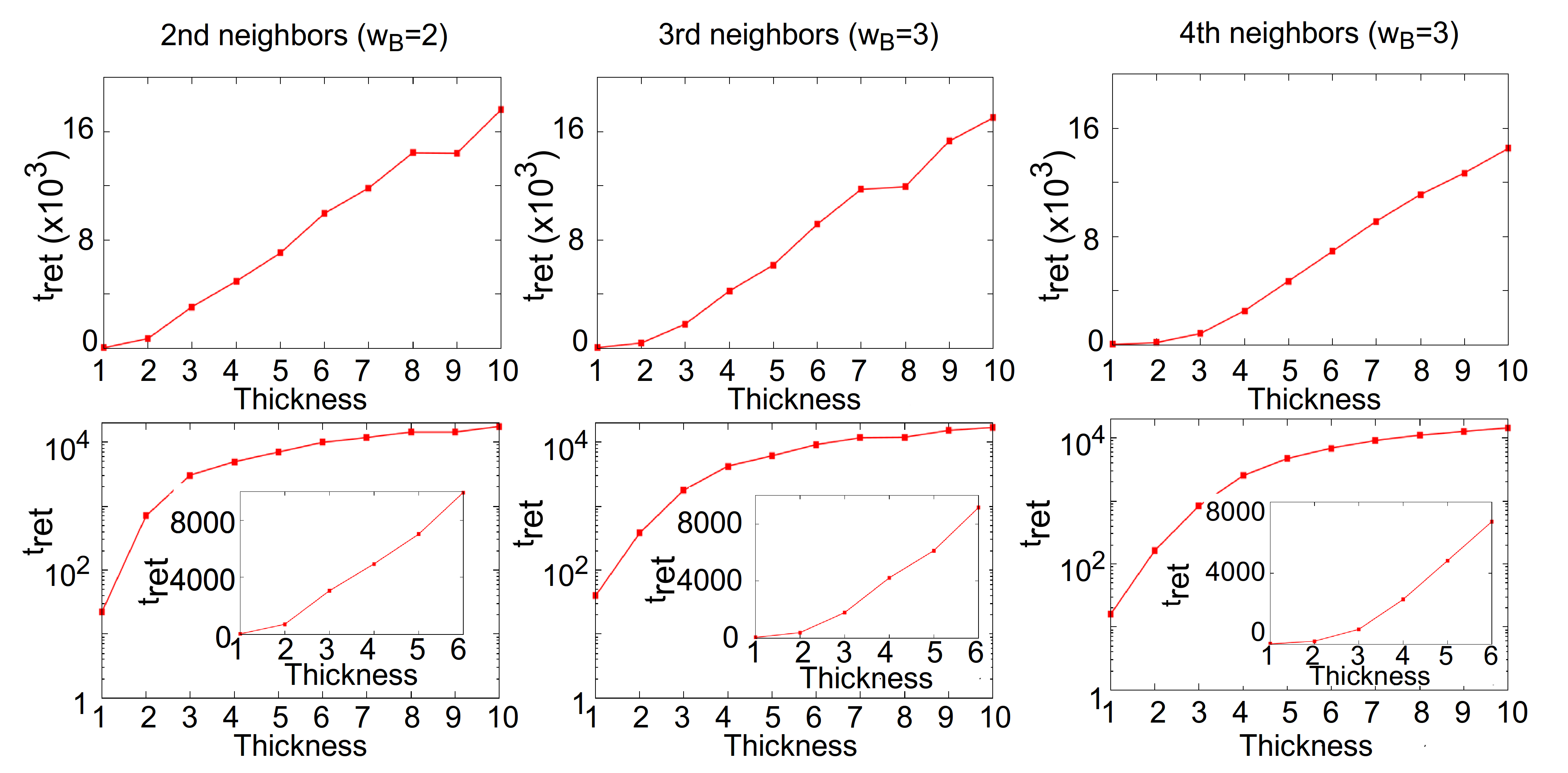}
  \caption{ Simulations for different interaction lengths. We show the results for 2nd, 3rd and 4th neighbors interactions. The plots show the variation of the retentivity as a function of cluster diameter. The two regimes explained in the text can be observed for $d<w_B$ and $d>w_B$. The results are shown in linear (top) and semi-log (bottom) scale. The inset in the bottom panels show the linear plots in a limited range for a better view of the change of regime.}
  
  \label{fig:longint}
\end{figure}

In this section we present simulations for longer interactions (beyond the n.n. used in maintext). In Fig.\ref{fig:longint} we show the results with interactions ranging from the 2nd to 4th layer of neighbors. In this case we can see that the evolution of $\tau_R$ as a function of the thickness exhibits first an exponential regime whose extension depends on the corresponding $w_B$. After this the expected linear behavior can be seen as explained in maintext and in this supplement. To keep the variation of the energy (and so $\tau_R$) in the same range of magnitude for every interaction (due to computation times), we used a different surface constant ($K$) for each case. We used $K=0.5$, $K=0.33$ and $K=0.2$ for the 2nd, 3rd and 4th neighbors interaction respectively.

\end{document}